\documentclass{aastex701}

\usepackage{amsmath}\usepackage{graphicx}
\usepackage{subcaption} 
\usepackage{booktabs} 
\usepackage{array} 

\begin{document}
\title{Fast radio bursts track cosmic star formation history with time delay: the effect of burst fluence}

\author[0009-0009-3583-552X]{X. D. Jia}
\affiliation{School of Astronomy and Space Science, Nanjing University, Nanjing 210093, China}
\email{602023260009@smail.nju.edu.cn}

\author[0000-0001-7176-8170]{D. H. Zhai}
\affiliation{School of Astronomy and Space Science, Nanjing University, Nanjing 210093, China}
\email{2300761756@qq.com}

\author[0000-0002-5819-5002]{J. P. Hu}
\affiliation{Ministry of Education Key Laboratory for Nonequilibrium Synthesis and Modulation of Condensed Matter, School of Physics, Xi’an Jiaotong University, Xi’an 710049, China}
\email{972143860@qq.com}

\author[0000-0003-0672-5646]{Shuang-Xi Yi}
\affiliation{School of Physics and Physical Engineering, Qufu Normal University, Qufu 273165, People's Republic of China}\email{yisx2015@qfnu.edu.cn}

\author[0000-0003-4157-7714]{F. Y. Wang}
\affiliation{School of Astronomy and Space Science, Nanjing University, Nanjing 210093, China}
\affiliation{Key Laboratory of Modern Astronomy and Astrophysics (Nanjing University), Ministry of Education, China}
\email{fayinwang@nju.edu.cn}

\correspondingauthor{F. Y. Wang}
\email{fayinwang@nju.edu.cn}

\begin{abstract}
Fast radio bursts (FRBs) are luminous millisecond-duration radio transients of extragalactic origin, yet their progenitor systems remain unidentified. The redshift evolution of the FRB event rate provides a powerful, model-independent discriminant between young progenitors and old progenitors, but recent studies have reached conflicting conclusions. In this work, we perform a forward-modeling analysis of the apparent non-repeating FRB sample from the CHIME/FRB catalogs to constrain the intrinsic FRB population. We find that the FRB rate follows the star formation rate (SFR) with little time delay when both dispersion measure (DM) and fluence are considered. In contrast, when only DM is taken into account, a significant time delay ($3 \sim 4$ Gyr) exists between the FRB rate and the SFR. The main reason for this discrepancy is that systematic biases exist in the mapping process from burst fluence to signal-to-noise ratio. These results support that at least a fraction of FRBs arise from a progenitor population whose event rate is delayed relative to star formation.
\end{abstract}


\section{Introduction} 
Fast radio bursts (FRBs) are millisecond-duration, highly dispersed flashes of radio emission, most of which originate at extragalactic and cosmological distances \citep{lorimer2007, thornton2013}. To date, with the improvement of observational techniques, more than four thousand FRB sources have been reported \citep{CHIME_catalog1,CHIME_catalog2}. Despite the fact that the physical mechanisms behind their huge isotropic energies are still unknown, FRBs are widely considered to originate outside the Milky Way due to their high repetition rates and large dispersion measure (DM) \citep{petroff2022, zhang2023, Wang2026,2026arXiv260925810S}. 

FRBs are observed in diverse environments, and their progenitor systems have not yet been conclusively identified. The detection of a bright radio burst from the Galactic magnetar SGR~1935+2154 \citep{2020Natur.587...54C, 2020Natur.587...59B} established magnetars formed in core-collapse supernovae as a viable engine, pointing toward a young stellar progenitor population.  By contrast, the repeating source FRB 20200120E was localized to a globular cluster of the nearby galaxy M81 \citep{2021ApJ...910L..18B, 2022Natur.602..585K}, indicating that at least some FRBs arise from old stellar environments. Thus, the proposed formation channels can be broadly categorized into two evolutionary pathways characterized by different cosmological histories \citep{zhang2023}: (i) A prompt channel, which traces the cosmic star formation history (SFR) with negligible delay, such as magnetars born in core-collapse supernovae or supermassive neutron stars that collapse into black holes \citep{2014A&A...562A.137F, 2014MNRAS.441.2433R, 2014ApJ...780L..21Z}; (ii) A delayed channel, driven by compact binary mergers, in which the event rate is obtained by convolving the star formation history with a delay-time distribution (DTD), such as double neutron star mergers, magnetars formed through mergers, or accretion-induced collapse of white dwarfs \citep{Wang2016,2018ApJ...858...89C, 2019ApJ...886..110M,Wang2020}. Because these two channels predict distinct redshift evolutions of the event rate, FRB rate provides a powerful and largely model-independent discriminator of progenitor scenarios.

The direct way to constrain the FRB event rate and its dependence on redshift and energy is through a population study based on a large sample of bursts. The Canadian Hydrogen Intensity Mapping Experiment (CHIME) \citep{2018ApJ...863...48C} has provided a large volume of FRB observations with a single telescope. However, although these bursts share a common instrumental origin, the redshift and energy distributions inferred from this dataset using different statistical methods remain controversial. On one hand, several studies suggest that the FRB population traces the SFR. Population synthesis analyses have also provided strong evidence that the FRB rate evolves with the SFR, even allowing for a short ($0.1 - 1$ Gyr) delay time \citep{2024A&A...690A.377W}. Using forward modeling and Bayesian hierarchical analysis, \cite{wangyy2026} found that the peak of the FRB event rate essentially coincides with the peak of the SFR. On the other hand, some studies indicate that the FRB population is delayed with respect to the SFR \citep{2019MNRAS.487.3672Z, 2022MNRAS.511.1961H, 2022JCAP...01..040Q, 2022ApJ...924L..14Z, 2025ApJ...988L..64C, 2025ApJ...986..100G}. \cite{2019MNRAS.487.3672Z} analyzed FRB samples from Parkes and Australian Square Kilometer Array Pathfinder (ASKAP) to study the energy function and formation rate, incorporating metallicity effects. Their results indicate a moderate delay of $3 \sim 5$ Gyr between the FRB formation rate and the SFR. The same delay between the FRB rate and the SFR has also been derived using a backward non-parametric method under the Lynden-Bell framework \citep{Chenjh2024, 2025A&A...698A..18Z, 2025ApJ...988L..64C, Jia2026, 2026arXiv260704792Z}. Meanwhile, some studies take a more neutral stance, suggesting that current data can not well constrain the time delay between the FRB event rate and the SFR \citep{Shin2023,2024ApJ...962...73L}.

The differing conclusions drawn from various methods have sparked extensive discussions on the nature of the FRB population. In this work, we adopt a forward modeling approach and thoroughly examine how parameter choices and model assumptions affect the results. We further compare our findings with previous results obtained from backward modeling \citep{Jia2026}, aiming to probe the intrinsic properties of the FRB population. We investigate the relationship between the FRB event and the SFR by constructing a two-dimensional binned map of DM and the signal-to-noise ratio (S/N). Using a forward modeling approach, we analyze the FRB sample from CHIME while accounting for the effects of selection biases. The structure of this paper is as follows. In Section \ref{Sec:Method}, we describe our forward-modeling framework, including the two-dimensional binning scheme, the rate-evolution models, and the fluence and DM distributions. In Section \ref{Sec:data}, we describe the data sample and the selection criteria. In Section \ref{Sec:Result}, we present the results for the CHIME/FRB samples. Finally, we summarize our main conclusions in Section \ref{Sec:Conclusion}.

\section{Method} \label{Sec:Method}

\subsection{Two-dimensional binning scheme}
The fluence and DM of the FRBs encode information about their population properties. The model described in detail by \cite{2022MNRAS.509.4775J} includes componenets that specify how both fluence and DM depend on redshift. This population model encodes information about FRB energetics, the possible cosmic-time evolution of the FRB population, and the distance distribution inferred from DM observations. 

The joint distribution of fluence ($F$) and DM can be expressed as
\begin{equation}
    P(\lambda_{\mathrm{source}} \, | \, \lambda_{\mathrm{pop}}) = P(F, DM \, | \, \lambda_{\mathrm{pop}}) P(z \, | \, \lambda_{\mathrm{pop}}).
\end{equation}
Here, $\lambda_{\mathrm{source}}$ represents parameters related to the intrinsic properties of individual FRB sources, such as $F$ and DM, whereas $\lambda_{\mathrm{pop}}$ denotes parameters characterizing the FRB population, including the time delay. However, the poor localization of FRBs makes it difficult to calibrate the beam attenuation, leading to a systematic underestimate of the measured fluence. This bias prevents the direct use of fluence measurements in population studies, as it shifts the inferred FRB event rate toward higher redshifts and consequently reduces the estimated time delay. 

To avoid the systematic bias arising from the measured fluence, we employ a mapping method based on two-dimensional parameter binning. By leveraging the correlation between fluence and S/N, this approach provides an elegant way to map the data without relying directly on the measured values. To compare the model predictions with the observed data, we construct a two-dimensional histogram of the observed S/N and DM. The S/N and DM axes are binned logarithmically into $N_{\mathrm{S/N}}=15$ bins spanning $\mathrm{S/N}\in[12,\,200]$ and $N_{\mathrm{DM}}=20$ bins spanning $\mathrm{DM}\in[100,\,3500]~\mathrm{pc~cm}^{-3}$, respectively \citep{Shin2023}. The observed number counts in each two-dimensional bin are denoted by $n_{ij}$, where $i$ indexes the S/N bin and $j$ indexes the DM bin.

The expected number of detections in each bin, $\zeta_{ij}$, is evaluated by reweighting the CHIME/FRB injection sample as 
\begin{equation}
    \zeta_{ij} = \sum_{m=1}^{n_{ij}} W_m(F, \mathrm{DM}, \Theta, \lambda_{\mathrm{pop}}),
\end{equation}
where the parameter vector $\Theta = (\tau, w)$. Each recovered injection is assigned a weight
\begin{equation}
    W_m = \frac{\Delta T \, f_{\mathrm{sky}}}{N_{\mathrm{inj}} \, \varepsilon_{\mathrm{inj}}} \; \frac{P(F, \mathrm{DM} \, | \, \lambda_{\mathrm{pop}})}{P_{\mathrm{init}}(F, \mathrm{DM})} \; P_{\Theta}(F, \mathrm{DM}),
\end{equation}
where $\Delta T$ is the total survey exposure time, $f_{\mathrm{sky}}$ is the fraction of the sky covered by the survey, $N_{\mathrm{inj}}$ is the total number of injected events, and $\varepsilon_{\mathrm{inj}}$ is the injection efficiency. The term $P(F, \mathrm{DM} \, | \, \lambda_{\mathrm{pop}})$ is the model prediction for the FRB rate density evaluated at the fluence and DM of the $m$-th recovered injection, and $P_{\mathrm{init}}(F_m, \mathrm{DM}_m)$ is the distribution from which the injections were drawn. The ratio $P/P_{\mathrm{init}}$ therefore reweights the injected events from their initial distribution to the model distribution, while $P_{\Theta}$ accounts for the reweighting of the pulse-width and scattering-time distributions from their injected forms to the fiducial distributions. The expected counts are obtained by summing the weights of all recovered injections falling into each bin. The likelihood is then the product of independent Poisson distributions over all bins
\begin{equation}
    \ln \mathcal{L} \propto \sum_{i,j} \left[ n_{ij} \ln \zeta_{ij} - \zeta_{ij} \right].
\end{equation}

The adopted priors and
model details are summarized in the Appendix.

\section{Data} \label{Sec:data}
\begin{figure}[h]
    \centering
    \includegraphics[width = \linewidth]{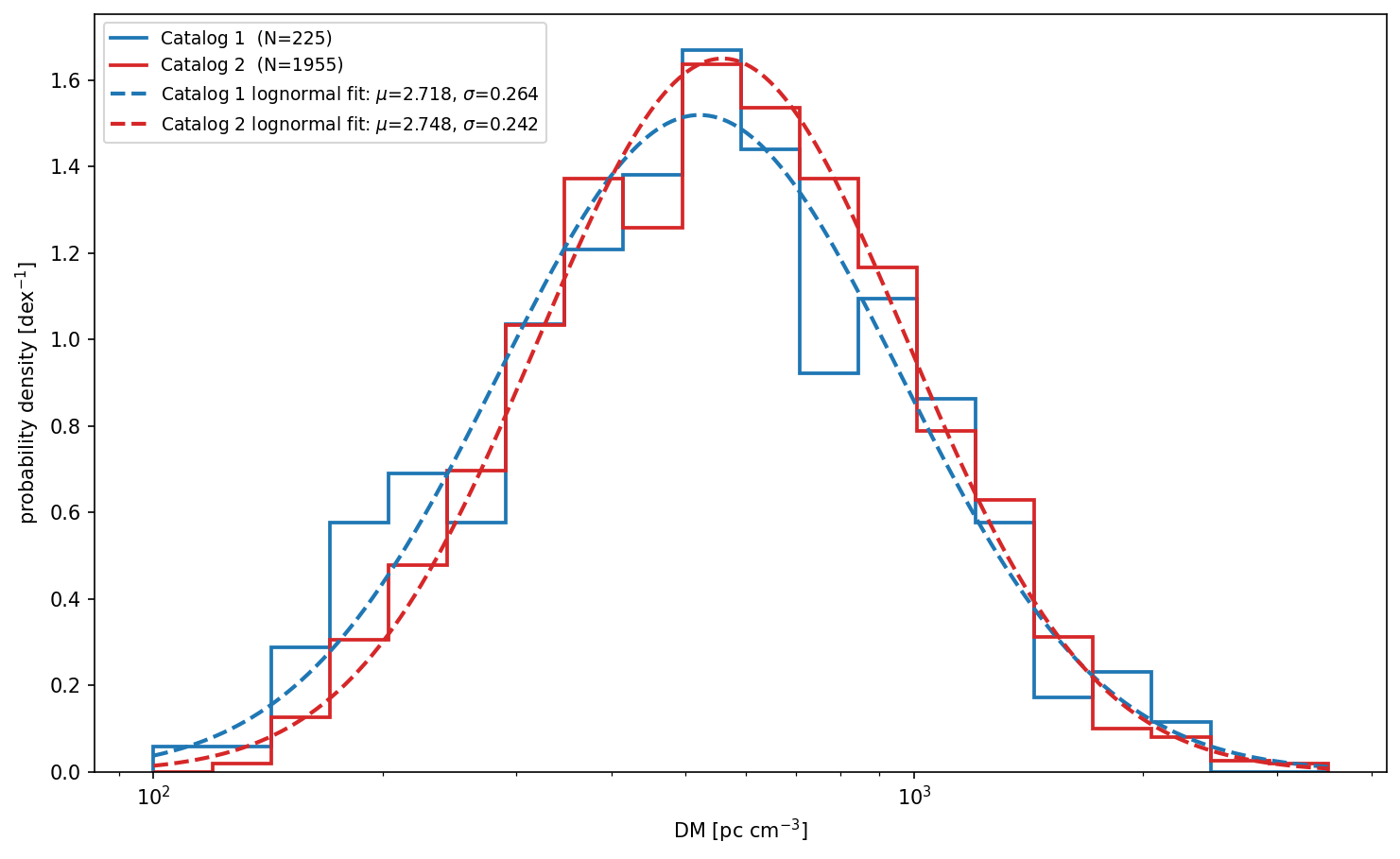}
    \caption{The DM distribution of Catalog 1 and Catalog 2. Both distributions are normalized, with blue and red representing Catalog 1 and Catalog 2, respectively. The dashed lines show the results of log-normal fits. The KS test yields a value of $p = 0.26$.}
    \label{fig:DM distribution}
\end{figure}

CHIME has discovered thousands of FRBs since $2018$ \citep{CHIME_catalog1,CHIME_catalog2}. The latest released CHIME/FRB catalog contains $4539$ FRB events, spanning from $2018$ July to $2023$ September, with a total exposure of $1837$ days. The Catalog comprises $4539$ bursts, including apparently $3558$ one-off bursts and $981$ repeating bursts. In this paper, we primarily focus on the $3558$ one-off bursts. Although this sample constitutes a large FRB catalog from a single survey with uniform selection effects, a subset of events may still be affected by certaion selection effects that are difficult to quantify. Therefore, in order to model the intrinsic property distribution of FRBs, we restrict our analysis to the subset of events whose statistical properties are most robust. To this end, we adopt the following criteria.
\begin{enumerate}
    \item To mitigate the tisk of false positives arising from human identification, FRB candidates with S/N $< 12$ are removed from the sample.

    \item To avoid misclassifying extragalactic bursts as Galactic ones, we excluded events with DM $<$ $1.5$ max(DM$_{\mathrm{NE2001}}$, DM$_{\mathrm{YMW16}}$).

    \item Since we adopts the same assumption as \cite{Shin2023}, namely DM$_{\mathrm{MW}} = 80$ pc cm$^{-3}$, we excluded events with DM $<$ $100$ pc cm$^{-3}$ to avoid unphysical negative values of DM$_{\mathrm{ex}}$.

    \item To mitigate the large uncertainties of the Galactic DM contribution near the Galactic plane, we remove FRBs with DM$_\mathrm{MW}$ $>$ $200$ pc cm$^{-3}$ \citep{2026PhRvL.137d1001W}.

    \item To avoid selection biases arising from CHIME istrumental effects, we excluded highly scattered events ($\tau > 100$ms at $600$ MHz).
\end{enumerate}
Following these selections, $1955$ events remain in Catalog 2, forming the basis of our analysis. For Catalog 1, we use the sample obtained after the selection criteria of \cite{Shin2023}. In Figure \ref{fig:DM distribution}, we show the distribution of DM for the two samples. A Kolmogorov-Smirnov (KS) test yields a $p$ value of $p = 0.26$, indicating that there is insufficient evidence to reject the null hypothesis that both samples are drawn from the same distribution. The normalized distributions are shown in blue (Catalog 1) and red (Catalog 2). The dashed lines represent the log-normal fitting results. As shown in Figure \ref{fig:DM distribution}, the peak of Catalog 2 is shifted to the right relative to that of Catalog 1, located at approximately 450 $~\mathrm{pc~cm}^{-3}$ compared to $\sim$350 $~\mathrm{pc~cm}^{-3}$ for Catalog 1. Moreover, Catalog 2 has a significantly larger high-DM tail than Catalog 1.

The injected data used in this work are taken from a synthetic injection system \citep{2023AJ....165..152M}. It is designed to properly account for instrumental effects by injecting synthetically generated signals into the software detection pipeline, thereby capturing the impact of real-time detection processes. Further details can be found in \cite{2021ApJS..257...59C}. Because the injected data are synthesized according to the instrument characteristics, and the distributions of Catalog 1 and 2 are similar, it is reasonable to combine the injection data with the sample to infer the intrinsic properties of FRBs.

\section{Result} \label{Sec:Result}
We first validate our method by adopting the same parameterization as \cite{Shin2023}, and obtain a result consistent with theirs within the $1\sigma$ level. 
Due to the lack of localized FRBs, the constraint on the index of the SFR $n$ ($(\frac{\mathrm{SFR(z)}}{\mathrm{SFR(0)}})^n$) is not precise. However, within the $1\sigma$ range, the value of $n$ is consistent with $1$. The intrinsic energy distribution is described by a power-law slope $\gamma \simeq -1.6$ and a characteristic cutoff energy $\log_{10}(E_{\mathrm{char}}/\mathrm{erg}) \simeq 41.6$. The spectral index is $\alpha = -1.33^{+0.66}_{-0.91}$, which is consistent with the result obtained by \cite{Shin2023}. However, it differs significantly from that derived under the delayed channel, which will be discussed in detail later.

\subsection{Delay time distribution}
\begin{figure}
    \centering
    \includegraphics[width=\linewidth]{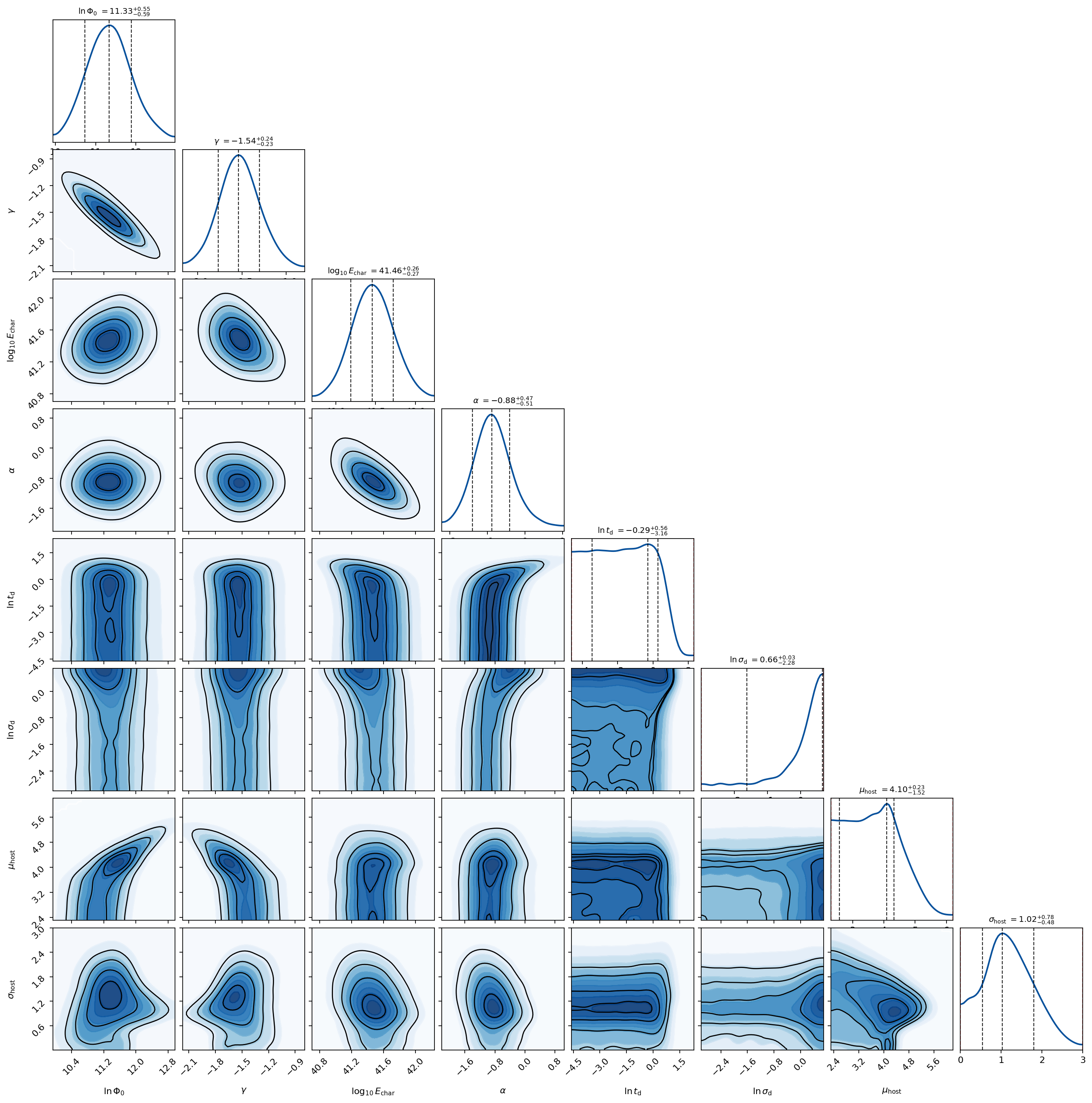}
    \caption{The corner plot of Catalog 1 from S/N and DM. The panels on the diagonal show the 1D histogram for each parameter obtained by marginalizing over the other parameters. The off-diagonal panels show two-dimensional projections of the posterior probability distributions for each pair of parameters, with contours to indicate $1\sigma-3\sigma$ confidence levels.} \label{fig:corner_catalog1_delay}
\end{figure}

\begin{figure}
    \centering
    \includegraphics[width=\linewidth]{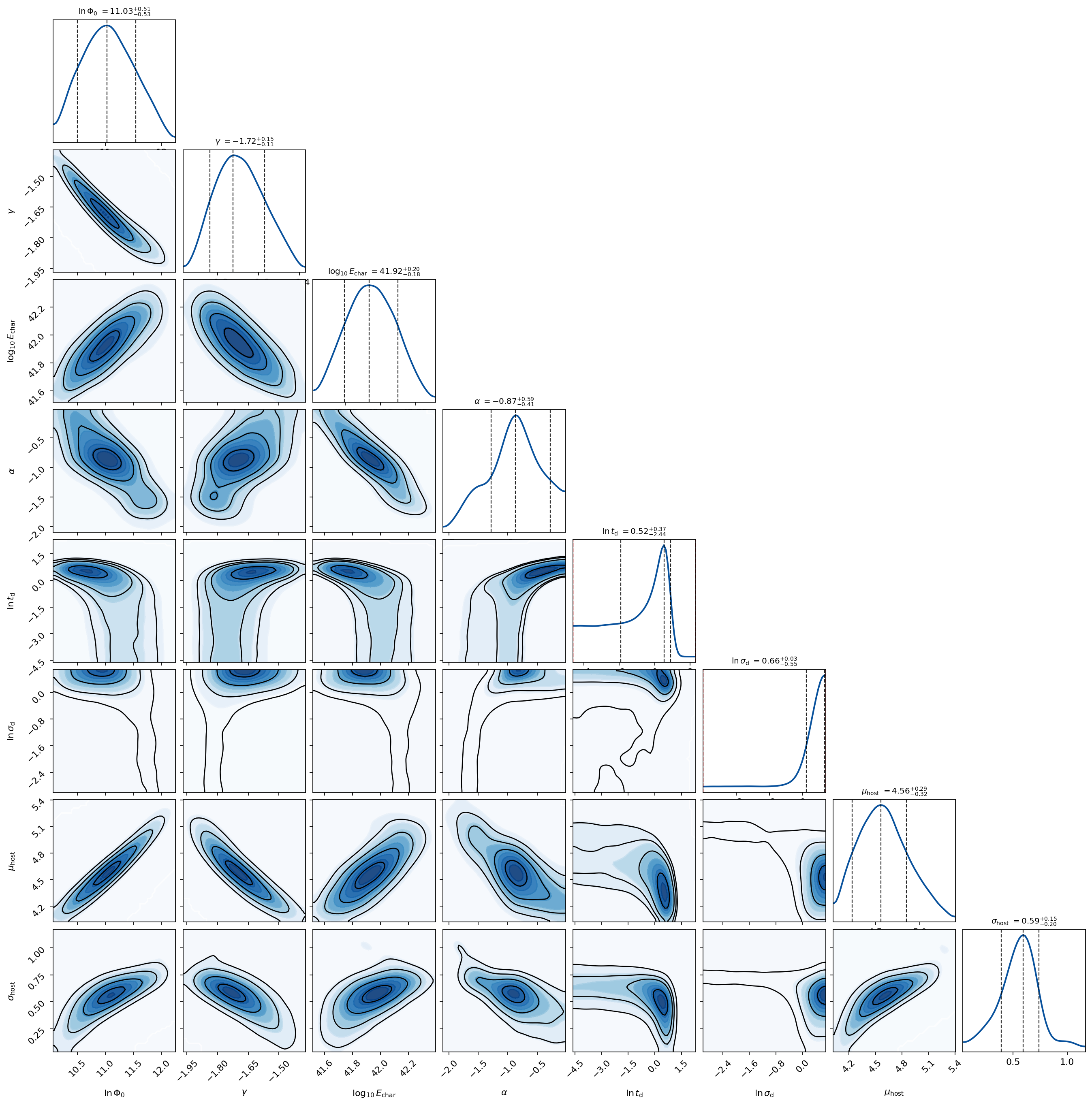}
    \caption{Same as Figure \ref{fig:corner_catalog1_delay}, but for Catalog 2.}
    \label{fig:corner_catalog2_delay}
\end{figure}

\begin{figure}[htbp]
    \centering
    
    \begin{subfigure}[b]{0.8\textwidth} 
        \centering
        \includegraphics[width=\linewidth]{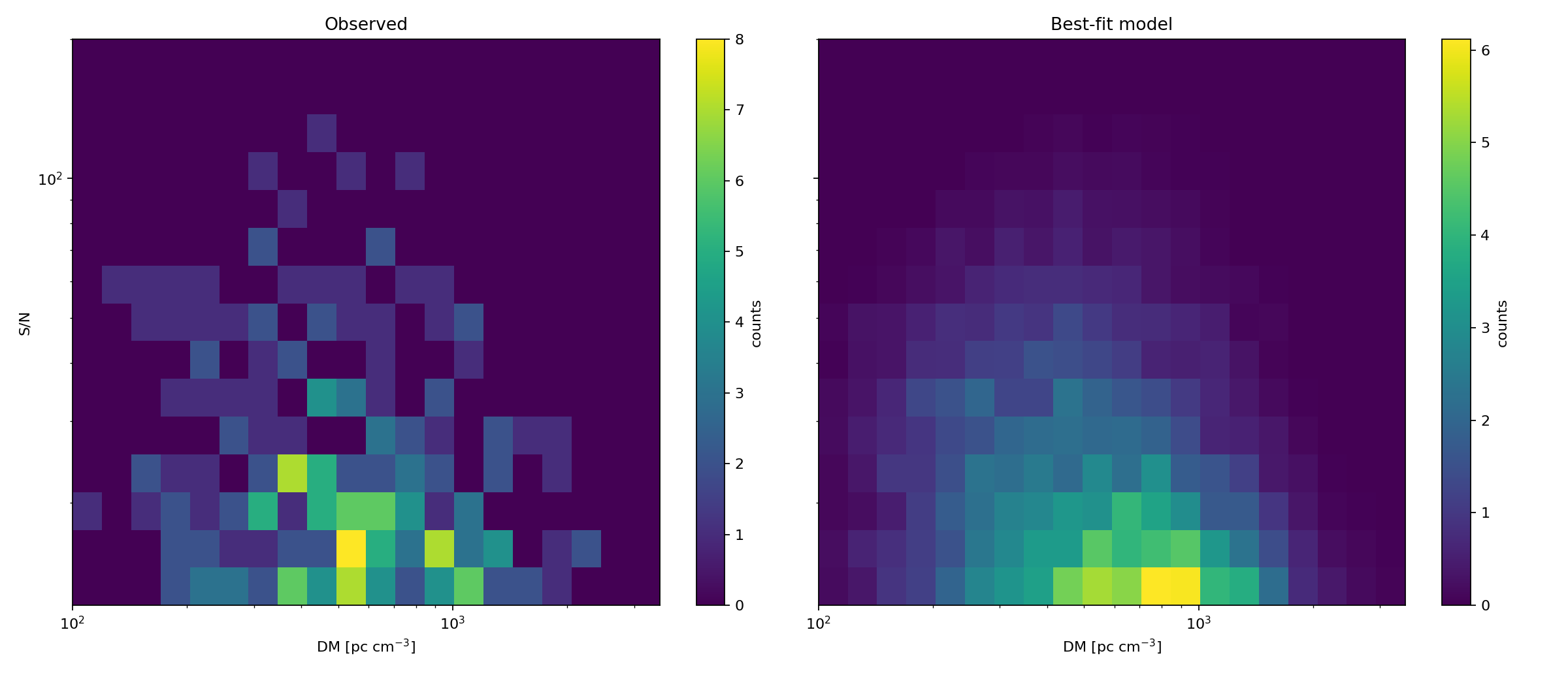}
        \caption{Fitting results for Catalog 1. The best-fit values are taken as the means of the posterior distributions. Both panels exhibit similar characteristics.}
        \label{fig:frb_obs_vs_model_prompt_catalog1}
    \end{subfigure}
    
    \vspace{0.5cm} 
    
    \begin{subfigure}[b]{0.8\textwidth}
        \centering
        \includegraphics[width=\linewidth]{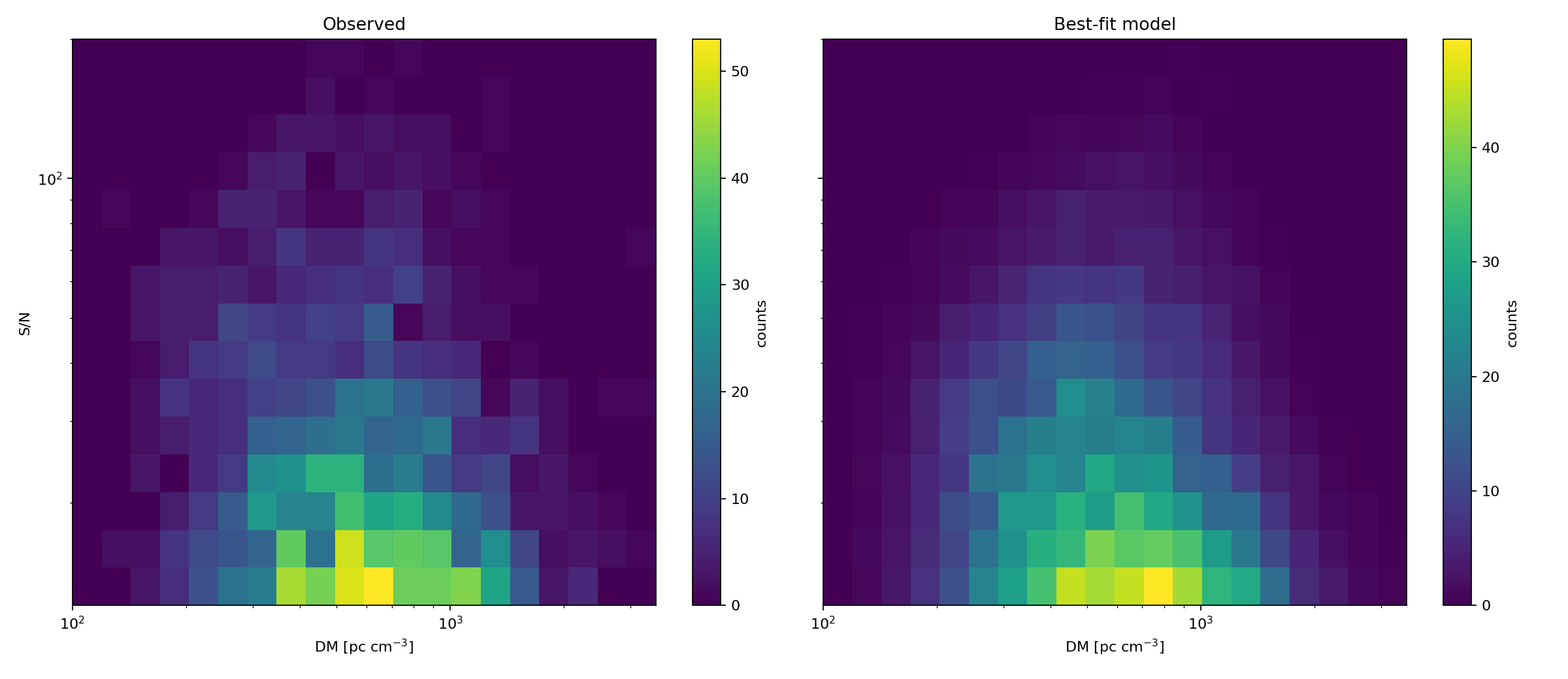} 
        \caption{Same as panel (a) but for Catalog 2.}
        \label{fig:frb_obs_vs_model_prompt_catalog2}
    \end{subfigure}
    
    \caption{Comparison of the S/N-DM distribution between the observed data (left) and the best-fit model prediction (right).}
    \label{fig:frb_obs_vs_model_prompt}
\end{figure}

We extend our analysis to the delayed channel, fitting Catalog 1 and 2 separately. The posterior distribution is sampled with the affine-invariant MCMC ensemble sampler \texttt{emcee} \citep{2013PASP..125..306F}, using $64$ walkers and $50{,}000$ steps, of which the first $20{,}000$ are discarded as burn-in. The parameters to be fitted in the current model include the following eight: $\{\Phi_0, \gamma, E_{\mathrm{char}}, \alpha, t_d, \sigma_{t_d}, \mu_{\mathrm{host}}, \sigma_{\mathrm{host}}\}$. When both DM and S/N are taken into account, the corner plot of the fitting results for Catalog 1 and 2 are shown in Figures \ref{fig:corner_catalog1_delay} and \ref{fig:corner_catalog2_delay}. The best-fit parameters are presented in the third and fifth columns of Table \ref{tab}. The comparison of the S/N-DM distribution between the observed data and the best-fit model prediction are shown in Figure \ref{fig:frb_obs_vs_model_prompt}.

Both Catalogs yield a small time delay, which is consistent with zero within $2\sigma$. Such results are not enough to suggest that the FRB rate is delayed relative to the SFR. The best-fit $\mathrm{DM}_{\mathrm{host}}$ increases and $\sigma_{\mathrm{host}}$ decreases, which are consistent with those derived from the IllustrisTNG simulation \citep{2020ApJ...900..170Z} and localized FRBs \citep{2025MNRAS.544.2870S}. It is worth noting that the results for $\mathrm{DM}_{\mathrm{host}}$ differ between Catalogs 1 and 2. This discrepancy is likely related to the differences in their DM distributions. 

The change in $\alpha$ is particularly noteworthy. In the forward modeling, parameters are degenerate due to the simultaneous fitting of multiple parameters. When the model is insufficient to adequately describe the intrinsic distribution represented by the data, it can lead to biases in one or several parameters. The spectral index $\alpha$ affects the redshift distribution of the data. We examined the effect of varying $\alpha$ while keeping all other parameters fixed: as $\alpha$ increases, the peak of the redshift distribution shifts toward lower redshifts. A similar discussion can be found in \cite{zhang2026}. Therefore, a possible explanation for this shift is as follows: if the intrinsic data distribution exhibits a delay but we fit it with a model that does not incorporate such a delay, the fitting procedure will inevitably suppress $\alpha$ in order to align a low-redshift-peaked dataset with a high-redshift-peaked model, yielding a fitted value smaller than the true one. We investigate this possibility in more detail in the following section.

\subsection{The Effect of S/N}
During the fitting process, we simultaneously use both DM and S/N to constrain the parameters. In this section, we discuss the impact of S/N on the results. 

Before employing the forward modeling with S/N and DM, we attempted a direct fit using $F$ and DM detected by CHIME. The result indicates that the FRB event rate follows the SFR. However, the $F$ measured by CHIME represent only the lower limits for a burst to be detected, rather than the true values. Consequently, underestimating $F$ leads to inferred distances larger than the true ones, thereby shifting the event rate toward higher redshifts. We then repeated the fit using the observed S/N and DM, and obtained similar results. This leads us to suspect that a similar issue may exist in the mapping between S/N and $F$. 

\begin{figure}
    \centering
    \includegraphics[width=\linewidth]{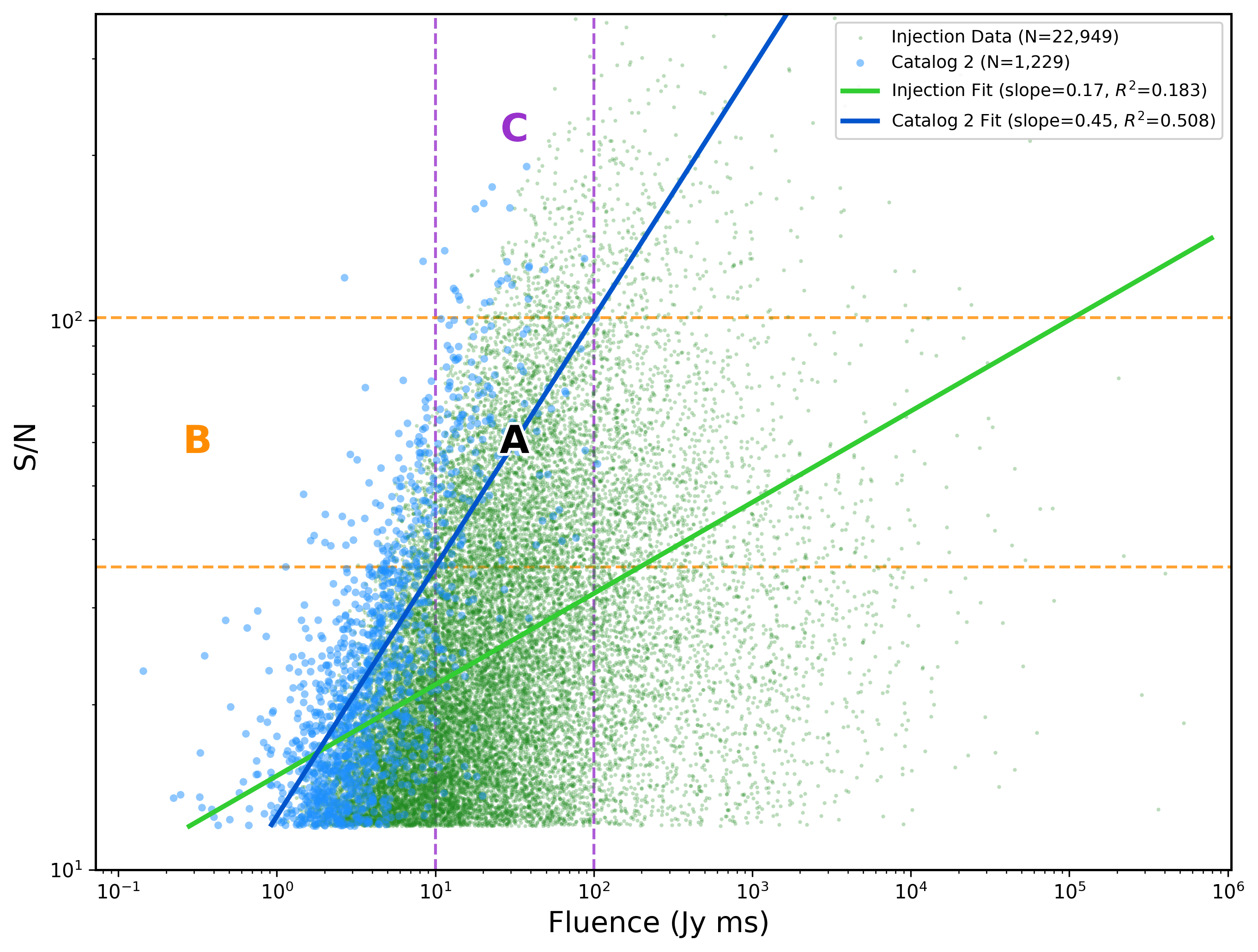}
    \caption{Distributions of S/N and DM for the injected data and Catalog 2. The blue points denote the data from Catalog 2, and the green points denote the injected data. The Catalog 2 points used for plotting are selected from the sample with an additional cut of $F \geq 0$, and are therefore fewer in number than those used in the text. The blue and green lines show their linear fits, respectively. The purple and orange dashed lines delineate two rectangular regions, B and C, whose overlapping region is denoted as A.}
    \label{fig:frb_distribution_c2+inj}
\end{figure}

Here we plot the distribution of S/N and $F$ in logarithmic coordinates, shown in Figure \ref{fig:frb_distribution_c2+inj}. From this figure, although S/N and $F$ are positively correlated, the relation is not strictly linear. This will lead to some systematic bias in the mapping process. In Figure \ref{fig:frb_distribution_c2+inj}, the blue points represent the data from Catalog 2, and the green points represent the injected data. Since plotting requires $F > 0$, the number of points shown in the figure is smaller than that actually used in the fitting. We draw two horizontal dashed lines and two vertical dashed lines, which divide the figure into rectangular regions B and C, together with their overlapping region A. For a well-behaved linearly correlated quantity, the data points should be concentrated around the fit line. In such a case, the number of data points in each S/N bin could be approximated by that in each $F$ bin, i.e., the number of points in region A. However, because the current data points are rather scattered, the actual number of points in each S/N bin corresponds to region B, while that in each $F$ bin corresponds to region C. Such a mapping inevitably introduces certain biases into the method, thereby affecting the posterior distributions of the fitted parameters.

\begin{figure}
    \includegraphics[width=0.5\linewidth]{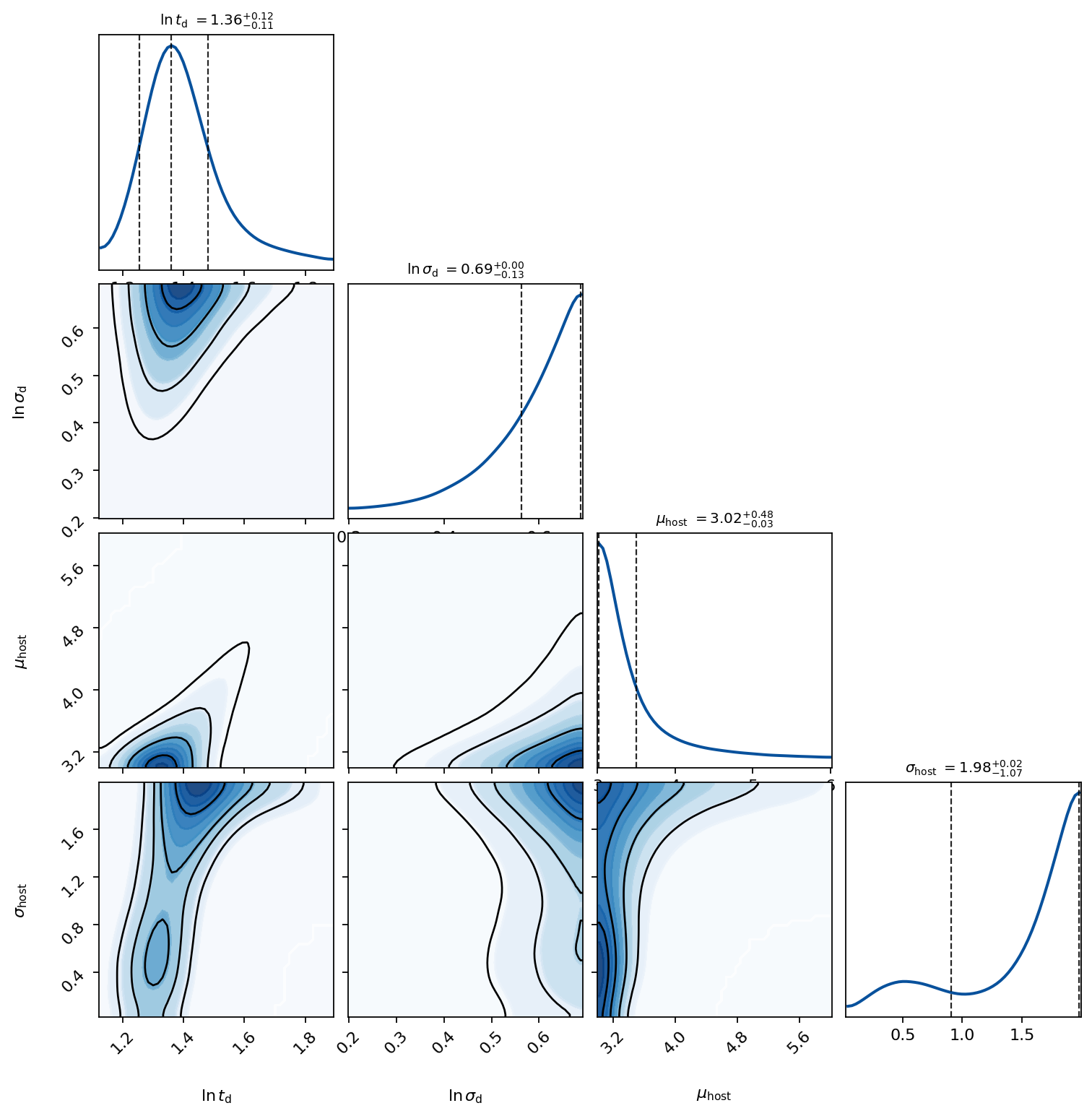}
     \includegraphics[width=0.5\linewidth]{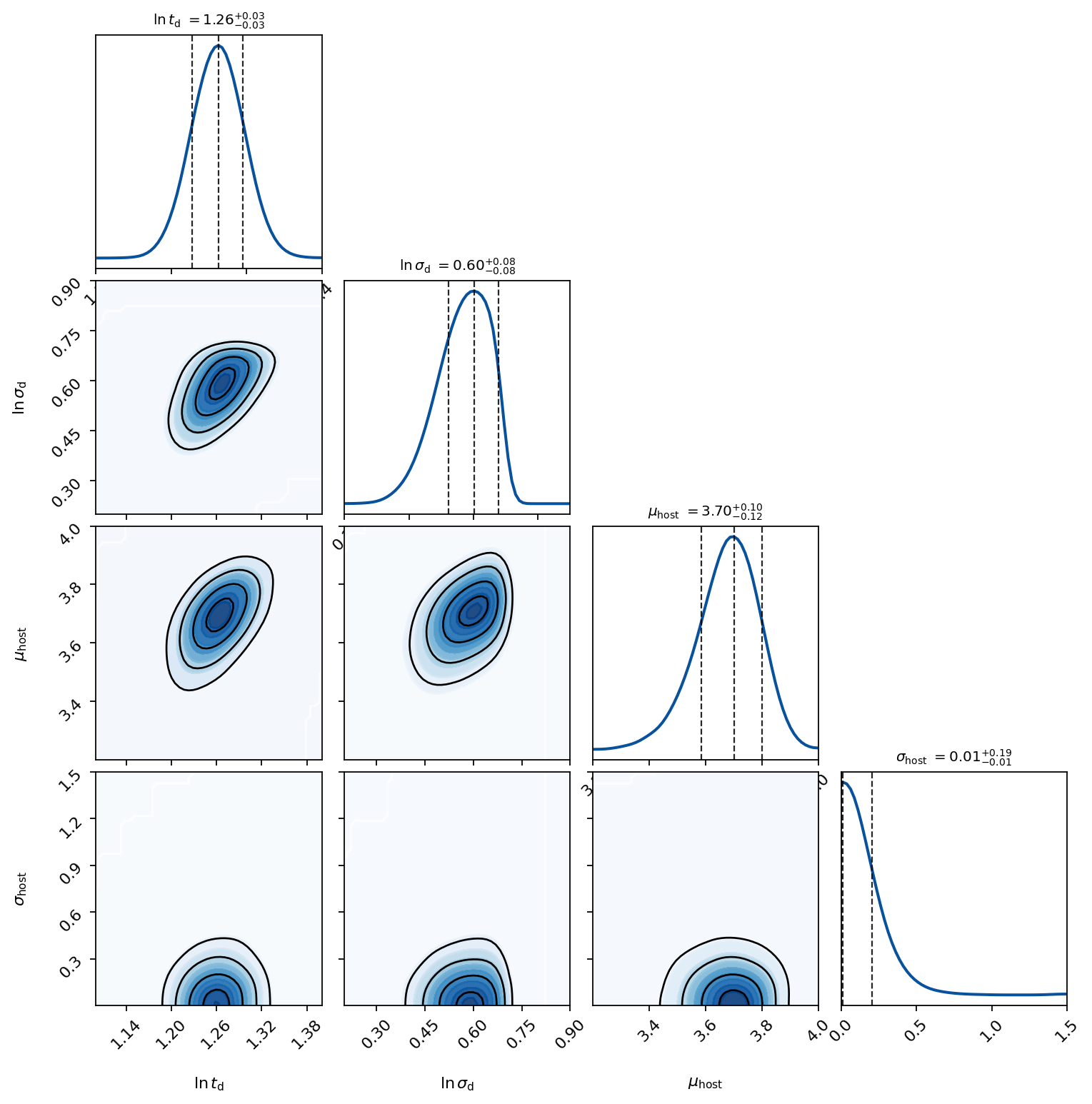}
    \caption{Same as Figure \ref{fig:corner_catalog1_delay}, but for Catalog 1 (left) and Catalog 2 (right) when only DM considered.}\label{fig:corner_Catalog_DM}
\end{figure}

Here, to avoid potential biases introduced by $F$, we investigate the distribution of the delay time considering only DM. The grouping strategy remains the same as described above, except that in this case we group only by DM and marginalize over fluence by integrating out the $F$-related probability. In this case, there are four parameters to be fitted $\{t_d, \sigma_{t_d}, \mu_{\mathrm{host}}, \sigma_{\mathrm{host}}\}$, and the best-fit values are presented in the fourth and sixth columns of Tables \ref{tab}. The corner plot of the fitting results are shown in Figure \ref{fig:corner_Catalog_DM}.

In contrast to the fitting results from S/N and DM, significant time delays are found for both Catalog 1 and Catalog 2 in the DM-only scenario. For Catalog 1, we obtain a delay time of $t_d = 3.90^{+0.50}_{-0.41}$ Gyr with $\sigma_{t_d} = 1.99^{+0.01}_{-0.24}$ Gyr. For Catalog 2, we obtain a delay time of $t_d = 3.53^{+0.11}_{-0.10}$ Gyr with $\sigma_{t_d} = 1.82^{+0.15}_{-0.14}$ Gyr. Both Catalogs 1 and 2 exhibit a significant time delay. The results are consistent with the time delay obtained via the Lynden-Bell method in backward modeling \citep{Chenjh2024, Jia2026, zhang2026}. This indicates that the treatment of fluence in the forward modeling process affects the inferred delay time.

\begin{figure}
    \centering
    \includegraphics[width=\linewidth]{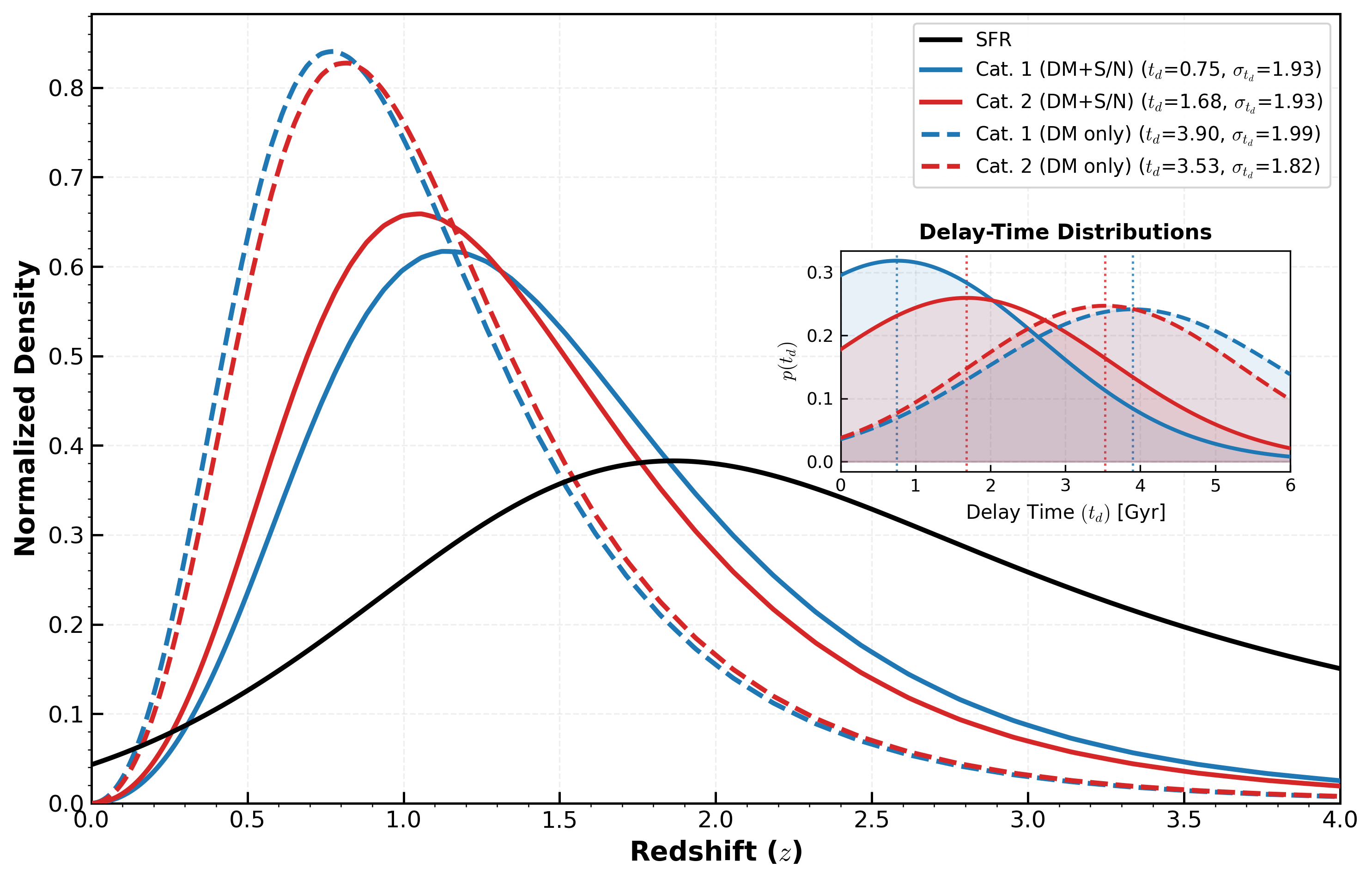}
    \caption{Delay time distributions of the FRB rate relative to the SFR for different samples and scenarios. The black line represents the SFR, the red lines represent the rate for Catalog 1, and the blue lines represent the rate for Catalog 2. The solid lines represent the joint use of DM and S/N, while the dashed lines correspond to using DM only.}
    \label{fig:pz}
\end{figure}

Figure \ref{fig:pz} gives the time delay distributions of the FRB rate relative to the SFR, obtained for different Catalogs in different scenarios. The black line represents the SFR, the red lines represent the rate for Catalog 1, and the blue lines represent the rate for Catalog 2. The main panel shows the relationship between the FRB rate and the SFR, while the right inset displays the delay time distribution.

Several systematic effects associated with the fluence $F$ merit careful consideration. The fluence $F$ reported by CHIME does not correspond to the true measured value, but the threshold fluence for triggering a detection. Adopting this value directly would substantially underestimate $F$, thereby biasing the inferred source distances toward larger redshifts and distorting the intrinsic redshift distribution. To mitigate this systematic bias, we adopt a two-dimensional binning scheme that exploits the empirical relation between $F$ and $\mathrm{S/N}$ \citep{Shin2023}. Nevertheless, an inspection of the joint distribution of $F$ and $\mathrm{S/N}$ in the injection samples (Figure~5 of \citealt{2023AJ....165..152M}) reveals that, although the two quantities are positively correlated, their relation is not strictly linear; as a consequence, the numbers of events mapped between corresponding bins in $F$ and $\mathrm{S/N}$ are not one-to-one. This implies that additional parameters—beyond the pulse width $w$ and scattering timescale $\tau$ already incorporated in the modeling—may modulate the $F$--$\mathrm{S/N}$ relation and potentially introduce residual systematic uncertainties into this correction. Furthermore, during the parameter inference, the fluence-related index $\alpha$ exerts a non-negligible influence on the inferred redshift distribution, as it directly affects the derived delay-time distribution. Finally, our fluence-based model does not account for a possible evolution of the burst energy with redshift. Although several studies have suggested that $E$ evolves with cosmic time \citep{Chenjh2024,Jia2026}, such an evolution has not yet been firmly established. We confine a tentative discussion of this possibility, leaving a more rigorous treatment to future work.

\subsection{The effect of DM}
For the probability distribution of $\mathrm{DM}_{\rm IGM}$, we adopt the functional form given in equation (\ref{p_IGM}). For the parameter values, we follow the parameterization of \cite{Shin2023}. However, we note that the parameters derived by \cite{2021ApJ...906...49Z} from IllustrisTNG simulations are consistent with them at $z > 0.5$, but differ significantly at $z < 0.5$. In this section, we discuss its effect on the delay time.

For equation (\ref{p_IGM}), we adopt two different parameter settings. In the first case, we set $\sigma_{\rm IGM} = 0.32z^{-0.5}$ \citep{2022MNRAS.509.4775J}, following the value used in \cite{Shin2023}. In the second case, we refer to the parameters derived from IllustrisTNG simulation in \cite{zhang2026} and obtain intermediate values via cubic spline interpolation. The parameters to be fitted in the current model include the following eight: $\{\Phi_0, \gamma, E_{\mathrm{char}}, \alpha, t_d, \sigma_{t_d}, \mu_{\mathrm{host}}, \sigma_{\mathrm{host}}\}$. The best-fit parameters of the two scenarios are consistent when both S/N and DM are considered, the delay time increases slightly to $2.1^{+1.1}_{-2.0}$ Gyr. The delay time is $t_d = 5.83^{+0.50}_{-0.37}$ Gyr when only DM is considered. We find that although the form of $\sigma_{\rm{IGM}}$ is changed, the conclusions remain unchanged in both cases. However, the resulting delay time is larger than that obtained with the previous form. 


We also consider the case where $\mathrm{DM}_{\rm host}$ evolves with redshift \citep{2020ApJ...900..170Z}. The parameterization is given by $\mathrm{DM}_{\rm host} = \mathrm{DM}_{\rm host,0} \, (1+z)^k$. Due to the degeneracy between $\mathrm{DM}_{\rm host,0}$ and $k$, which prevents simultaneous fitting, we fix $k$ as the value obtained in \cite{Mas-Ribas:2026uhy}. The fitting results show that for Catalog 2 with only DM considered, the delay time is $3.90^{+0.16}_{-0.15}$ Gyr with $\sigma_{t_d} = 1.92^{+0.08}_{-0.11}$ and $\mathrm{DM}_{\rm host,0} = 37^{+4.0}_{-4.2} $ pc cm$^{-3}$. The results are consistent with that obtained when $\mathrm{DM}_{\rm host}$ does not evolve with redshift, but the fit to the DM distribution is improved at high redshifts. 



\section{Conclusions} \label{Sec:Conclusion}
In this work, we have performed a forward-modeling analysis of CHIME/FRB catalogs to constrain the relationship between the FRB event rate and the cosmic SFR. 
We model the joint distribution of the S/N and DM through a two-dimensional binning scheme, and account for the survey selection function using the CHIME/FRB injection system. This approach jointly constrains the FRB rate evolution, the intrinsic energy distribution, and the host-galaxy DM contribution. 

\begin{enumerate}
    \item When both S/N and DM are considered, FRB rate follows SFR with little time delay. The delay time is consistent with zero in 2$\sigma$ confidence level.

        \item In the DM-only analysis, a clear time delay relative to the SFR is found in both CHIME/FRB Catalog 1 and 2, consistent with the backward-modeling results. This agreement underscores the need to account for further systematic effects in the $F$-to-S/N mapping.
        
    \item Mapping $F$ to S/N avoids the direct use of measured fluence values, but systematic biases inherent in this mapping remain to be addressed. Although S/N is positively correlated with $F$, it is also modulated by additional factors. These systematic effects may suppress the inferred delay of the FRB rate relative to the SFR, creating the impression that the FRB rate completely tracks the SFR.

    \item The effects of $\mathrm{\sigma}_{\rm IGM}$ and $\mathrm{DM}_{\rm host}$ on the inferred delay time are discussed. We found that, although some debate remains regarding the forms of $\mathrm{\sigma}_{\rm IGM}$ and $\mathrm{DM}_{\rm host}$, the differences between various models are not sufficient to significantly affect the time delay.
    
\end{enumerate}

Regarding the mapping from $F$ to $\mathrm{S/N}$, we acknowledge the presence of systematic biases and expect that the forthcoming localized FRBs releases will help to address this issue \citep{2026arXiv260626334M}. 
Upcoming facilities are expected to increase the number of localized FRBs to between $10^3$ and $10
^4$. These include the DSA-2000 \citep{Sherman2024}, the Hydrogen intensity and Real-time Analysis eXperiment \citep{Crichton2022HIRAX}, Canadian Hydrogen Observatory and Radio-transient Detector (CHORD) \citep{2019clrp.2020...28V} and the Square Kilometre Array \citep{Caleb2026}, which together will help to alleviate this degeneracy.

\begin{acknowledgments}
This work was supported by the National Natural Science Foundation of China (grant Nos. 12625304, 12494575 and 12273009).
\end{acknowledgments}

\begin{contribution}
X. D. Jia: Data curation, Software, Formal analysis, Methodology, Visualization, and Writing - original draft.\\
D. H. Zhai: Investigation and Writing - review and editing.\\
J. P. Hu: Writing - review and editing.\\
S. X. Yi: Writing - review and editing.\\
F. Y. Wang: Conceptualization, and Writing - review and editing. \\
All authors reviewed and approved the final version of the manuscript.
\end{contribution}

\appendix
\section{Time delay with respect to SFR}
For the redshift distribution of FRBs, we consider a delayed time channel. 
There is a evolutionary timescale between the formation of the progenitor system and the subsequent FRB event. The FRB rate can be written as the convolution of the SFR with a delay time distribution (DTD) \citep{2015MNRAS.448.3026W},
\begin{equation}
    \dot{R}(z\, | \, \lambda_{\mathrm{pop}}) = \Phi_0 \int_{0}^{t(z)-t_b} \mathrm{SFR}\left[t(z)-\tau\right] \, \mathrm{DTD}(\tau) \, d\tau,
\end{equation}
where $\tau$ is the delay time between the formation of the progenitor system and the FRB event, $\Phi_0$ is the local event-rate density normalization, and $t_b$ is the cosmic time at the onset of star formation. We adopt the parameterization of the SFR from \cite{2014ARA&A..52..415M}. For the DTD, we adopt a truncated Gaussian form,
\begin{equation}
    \mathrm{DTD}(\tau) \propto \exp\left[ -\frac{(\tau - t_d)^2}{2 \sigma_d^2} \right], \quad \tau > 0,
\end{equation}
characterized by a characteristic delay time $t_d$ and a width $\sigma_d$. In both scenarios, the probability density of the FRB redshift is given by
\begin{equation}
    p(z \, | \, \lambda_{\mathrm{pop}}) \propto \frac{\dot{R}(z \, | \, \lambda_{\mathrm{pop}})}{1+z} \, \frac{dV_c}{dz},
\end{equation}
where $dV_c/dz$ is the comoving volume element per unit redshift and the factor $(1+z)^{-1}$ accounts for cosmological time dilation.

\section{Fluence distribution}
We model the intrinsic energy distribution of FRBs with a Schechter function, characterized by a power-law slope $\gamma$ and an exponential cutoff at the characteristic energy $E_{\mathrm{char}}$ \citep{1976ApJ...203..297S}
\begin{equation}
    P(E) \propto \frac{1}{E_{\mathrm{char}}} \left( \frac{E}{E_{\mathrm{char}}} \right)^{\gamma} \exp\left( -\frac{E}{E_{\mathrm{char}}} \right), \quad E > E_{\mathrm{pivot}},
\end{equation}
where $E_{\mathrm{pivot}}$ is a pivot energy below which the distribution is not modeled. Such a form for the energy distribution of FRBs has also been considered by previous works \citep{2018MNRAS.481.2320L, 2021ApJ...909L...8N}. It is worth noting that the possible redshift evolution of $P(E)$ is not taken into account in this study \citep{Chenjh2024, Jia2026}. The isotropic-equivalent energy $E$ is related to the observed fluence $F$ through
\begin{equation}
    E = \frac{4\pi d_L^2(z) \, \Delta\nu}{(1+z)^{2+\alpha}} \, F,
\end{equation}
where $d_L(z)$ is the luminosity distance, $\Delta\nu$ is the observing bandwidth (taken as $1$ GHz), and $\alpha$ is the spectral index defined by $F_\nu \propto \nu^{\alpha}$. 
The conditional fluence distribution at a given redshift is then obtained by a change of variables,
\begin{equation}
    P(F, z) = P\left(E(F, z)\right) \, \frac{dE}{dF}.
\end{equation}
In our calculation, the luminosity distance is determined using the cosmological parameters form the results of Planck \citep{2020A&A...641A...6P}. Thus, the parameters related with fluence distribution are $\alpha$, the spectral index, $\gamma$ the power-law index, and $E_{\mathrm{char}}$ the exponential cutoff energy.

\section{DM distributions}
The observed DM serves as a fundamental observational quantity for FRBs. It is typically divided into the following four components, Milky Way (MW) interstellar medium (DM$_{\mathrm{MW}}$), Galactic halo (DM$_\mathrm{halo}$), intergalactic medium (IGM; DM$_{\mathrm{IGM}}$), and host galaxy (DM$_{\mathrm{host}}$):
\begin{equation}
    \mathrm{DM}_{\mathrm{obs}} = \mathrm{DM}_{\mathrm{MW}} + \mathrm{DM}_{\mathrm{halo}} + \mathrm{DM}_{\mathrm{IGM}} + \frac{\mathrm{DM}_{\mathrm{host}}}{1 + z}.
\end{equation}
Here, $\mathrm{DM}_{\mathrm{obs}}$ is the total observed DM and $\mathrm{DM}_{\mathrm{host}}$ is the DM of the host galaxy in the FRB's source frame, with the factor $(1+z)$ accounts for the time dilation. $\mathrm{DM}_{\mathrm{MW}}$, the DM contributed by the interstellar medium of the MW, is described by Galactic electron distribution models such as YMW16 and NE2001 \citep{2002astro.ph..7156C, 2017ApJ...835...29Y}. The component $\mathrm{DM}_{\mathrm{halo}}$ is currently poorly constrained. In this work, we adopt the same assumption for $\mathrm{DM}_{\mathrm{halo}}$ as in \cite{Shin2023}, setting $\mathrm{DM}_{\mathrm{halo}} = 30~\mathrm{pc~cm}^{-3}$ \citep{2015MNRAS.451.4277D, Shin2023}. Therefore, the contribution from the extragalactic components is commonly written as:
\begin{equation}
    \mathrm{DM}_{\mathrm{exc}} \equiv \mathrm{DM}_{\mathrm{obs}} - \mathrm{DM}_{\mathrm{MW}} - \mathrm{DM}_{\mathrm{halo}} = \mathrm{DM}_{\mathrm{IGM}} + \frac{\mathrm{DM}_{\mathrm{host}}}{1 + z}.
\end{equation}
It must be noted that the local environment can contribute a siginicant fraction of DM \citep{ 2022NatCo..13.4382W, 2026ApJ..1009..211W}.

Within the framework of the $\Lambda$CDM cosmological model, the mean value of $\mathrm{DM}_{\mathrm{IGM}}$ can be expressed as 
\begin{equation}
    \langle \mathrm{DM}_{\mathrm{IGM}} \rangle = \frac{3cH_0\Omega_b f_{\mathrm{IGM}}}{8\pi G m_p} \times f_e(z),
\end{equation}
where $c$ is the speed of light, $H_0$ is the Hubble constant, $\Omega_b$ is the densities of baryon matter, $f_{\mathrm{IGM}}$ is the fraction of baryons in the IGM, G is the Newtonian gravitational constant, $m_p$ is the proton mass, and $f_e(z)$ is defined as 
\begin{equation}
    f_e(z) = \int_0^z \frac{\left[ \frac{3}{4}y_1 X_{e,\mathrm{H}}(\tilde{z}) + \frac{1}{8}y_2 X_{e,\mathrm{He}}(\tilde{z}) \right] (1 + \tilde{z}) \, d\tilde{z}}{\left[ \Omega_m (1 + \tilde{z})^3 + \Omega_\Lambda \right]^{1/2}}.
\end{equation}
The parameters $y_1$ and $y_2$ are the hydrogen and helium fractions normalized to $0.76$ and $0.24$, respectively, which can be neglected as $y_1 \simeq y_2 \simeq 1$. In the late Universe ($z < 3$), hydrogen and helium can be considered fully ionized \citep{2009RvMP...81.1405M, 2011MNRAS.410.1096B}. Therefore, their ionization fractions $\chi_{e,\mathrm{H}}(z)$ and $\chi_{e,\mathrm{He}}(z)$ can be taken as $\chi_{e,\mathrm{H}}(z) = \chi_{e,\mathrm{He}}(z) = 1$. Finally, the expression for $\mathrm{DM}_{\mathrm{IGM}}$ can be simplified to
\begin{equation}
    \langle \mathrm{DM}_{\mathrm{IGM}} \rangle = \frac{21 c \Omega_b H_0^2}{64 \pi H_0 G m_p} \times \int_0^z \frac{f_{\mathrm{IGM}} (1+z)\,dz}{\left[ \Omega_m (1+z)^3 + 1 - \Omega_m \right]^{1/2}}.\label{DM_IGM}
\end{equation}
The value of $f_{\mathrm{IGM}} = 0.84$ is adopted \citep{2012ApJ...759...23S, Yang2022}. The cosmological parameters are adopted from the results of \cite{2020A&A...641A...6P}.

The Equation (\ref{DM_IGM}) gives the mean DM contribution from the IGM. For individual FRB events, the probability distribution of the true DM can be modeled by a Gaussian-like distribution as \citep{2020Natur.581..391M, 2021ApJ...906...49Z}
\begin{equation}
    p_{\mathrm{IGM}}(\Delta) = A \Delta^{-\beta} \exp\left[ -\frac{(\Delta^{-\alpha} - C_0)^2}{2\alpha^2 \sigma_{\mathrm{IGM}}^2} \right], \quad \Delta = \frac{\mathrm{DM}_{\mathrm{IGM}}}{\langle \mathrm{DM}_{\mathrm{IGM}} \rangle}.\label{p_IGM}
\end{equation}
The parameters $\alpha = \beta = 3$ \citep{2020Natur.581..391M}. The parameters $A$ and $C_0$ are normalization parameters chosen such that the distribution $p_{\mathrm{IGM}}$ is properly normalized ($\int p_{\mathrm{IGM}} = 1$) and $\langle \Delta \rangle = 1$.

Due to the lack of detailed observations of FRB host galaxies, the uncertainty in the host contribution to the DM is relatively large. Numerical simulations suggest that the probability distribution of DM follows a log-normal form \citep{2020Natur.581..391M, 2020ApJ...900..170Z},
\begin{equation}
    p_{\mathrm{host}}(\mathrm{DM}_{\mathrm{host}}) = \frac{1}{\sqrt{2\pi}\, \mathrm{DM}_{\mathrm{host}}\, \sigma_{\mathrm{host}}} \exp\left[ -\frac{(\ln \mathrm{DM}_{\mathrm{host}} - \mu)^2}{2\sigma_{\mathrm{host}}^2} \right],
\end{equation}
where $\mu$ and $\sigma_{\mathrm{host}}$ are the mean and standard deviation of ln$\mathrm{DM}_{\mathrm{host}}$, respectively. For the distribution of $\mathrm{DM}_{\mathrm{IGM}}$ in Equation (\ref{p_IGM}), we compared the results from the IllustrisTNG simulation \citep{2020ApJ...900..170Z, 2021ApJ...906...49Z} with the parameterization \citep{2022MNRAS.509.4775J}. The values of $C_0$ and $\sigma_\mathrm{IGM}$ derived from these two results are consistent at $z > 0.5$, but differ significantly at low redshifts. Here we adopt the same values as in \cite{Shin2023}, and discuss the impact of different choices on our results. Once the probability distributions of $\mathrm{DM}_{\mathrm{IGM}}$ and $\mathrm{DM}_{\mathrm{host}}$ are determined, the distribution of $\mathrm{DM}_{\mathrm{exc}}$ becomes a convolution:
\begin{equation}
    P(\mathrm{DM}_{\mathrm{exc}}, z) = \int d\mathrm{DM}_{\mathrm{host}} \, P(\mathrm{DM}_{\mathrm{host}}, z) \, P(\mathrm{DM}_{\mathrm{IGM}} = \mathrm{DM}_{\mathrm{exc}} - \mathrm{DM}_{\mathrm{host}}, z).
\end{equation}
Here, a change of variables is required in the integral. Therefore,
\begin{equation}
    P(\mathrm{DM}_{\mathrm{IGM}}, z) = P_{\mathrm{IGM}}(\Delta) \frac{d\Delta}{d\mathrm{DM}_{\mathrm{IGM}}} = P_{\mathrm{IGM}}(\Delta) \frac{1}{\langle \mathrm{DM}_{\mathrm{IGM}} \rangle (z)},
\end{equation}
and 
\begin{equation}
    P(\mathrm{DM}_{\mathrm{host}}, z) = P_{\mathrm{host}}(\mathrm{DM}'_{\mathrm{host}}) \frac{d\mathrm{DM}'_{\mathrm{host}}}{\mathrm{DM}_{\mathrm{host}}}.
\end{equation}
Here, $\mathrm{DM}_{\mathrm{host}}$ is in the observer frame, while $\mathrm{DM}_{\mathrm{host}}$ is in the source frame. Thus, the parameters related with DM are $\mathrm{DM}_{\mathrm{host}}$ and $\sigma_\mathrm{host}$.

%


\bibliography{sample701}{}

@ARTICLE{lorimer2007,
       author = {{Lorimer}, D.~R. and {Bailes}, M. and {McLaughlin}, M.~A. and {Narkevic}, D.~J. and {Crawford}, F.},
        title = "{A Bright Millisecond Radio Burst of Extragalactic Origin}",
      journal = {Science},
         year = 2007,
        month = nov,
       volume = {318},
       number = {5851},
        pages = {777},
          doi = {10.1126/science.1147532},
archivePrefix = {arXiv},
       eprint = {0709.4301},
 primaryClass = {astro-ph},
       adsurl = {https://ui.adsabs.harvard.edu/abs/2007Sci...318..777L}
}

@ARTICLE{thornton2013,
       author = {{Thornton}, D. and {Stappers}, B. and {Bailes}, M. and {Barsdell}, B. and {Bates}, S. and {Bhat}, N.~D.~R. and {Burgay}, M. and {Burke-Spolaor}, S. and {Champion}, D.~J. and {Coster}, P. and {D'Amico}, N. and {Jameson}, A. and {Johnston}, S. and {Keith}, M. and {Kramer}, M. and {Levin}, L. and {Milia}, S. and {Ng}, C. and {Possenti}, A. and {van Straten}, W.},
        title = "{A Population of Fast Radio Bursts at Cosmological Distances}",
      journal = {Science},
         year = 2013,
        month = jul,
       volume = {341},
       number = {6141},
        pages = {53-56},
          doi = {10.1126/science.1236789},
archivePrefix = {arXiv},
       eprint = {1307.1628},
 primaryClass = {astro-ph.HE},
       adsurl = {https://ui.adsabs.harvard.edu/abs/2013Sci...341...53T}
}

@ARTICLE{Wang2026,
       author = {{Wang}, Fayin and {Jia}, Xuandong and {Gao}, Daohong and {Dai}, Zigao},
        title = "{Fast Radio Burst Cosmology: Hubble Tension and Dark Energy}",
      journal = {Research in Astronomy and Astrophysics},
         year = 2026,
        month = aug,
       volume = {26},
       number = {8},
          eid = {084010},
        pages = {084010},
          doi = {10.1088/1674-4527/ae8428},
archivePrefix = {arXiv},
       eprint = {2606.22390},
 primaryClass = {astro-ph.CO},
       adsurl = {https://ui.adsabs.harvard.edu/abs/2026RAA....26h4010W}
}

@ARTICLE{Wang2020,
       author = {{Wang}, F.~Y. and {Wang}, Y.~Y. and {Yang}, Yuan-Pei and {Yu}, Y.~W. and {Zuo}, Z.~Y. and {Dai}, Z.~G.},
        title = "{Fast Radio Bursts from Activity of Neutron Stars Newborn in BNS Mergers: Offset, Birth Rate, and Observational Properties}",
      journal = {\apj},
         year = 2020,
        month = mar,
       volume = {891},
       number = {1},
          eid = {72},
        pages = {72},
          doi = {10.3847/1538-4357/ab74d0},
archivePrefix = {arXiv},
       eprint = {2002.03507},
 primaryClass = {astro-ph.HE},
       adsurl = {https://ui.adsabs.harvard.edu/abs/2020ApJ...891...72W}
}

@INPROCEEDINGS{Sherman2024,
       author = {{Sherman}, Myles and {Connor}, Liam and {Ravi}, Vikram and {Law}, Casey and {DSA-2000 Collaboration}},
        title = "{The DSA-2000 Fast Time Domain Search for Pulsars and Fast Radio Bursts}",
    booktitle = {American Astronomical Society Meeting Abstracts \#243},
         year = 2024,
       series = {American Astronomical Society Meeting Abstracts},
       volume = {243},
        month = feb,
          eid = {261.04},
        pages = {261.04},
       adsurl = {https://scixplorer.org/abs/2024AAS...24326104S}
}

@INCOLLECTION{Caleb2026,
       author = {{Caleb}, Manisha and {Fortunato}, J{\'e}ferson A.~S. and {Hagstotz}, Steffen and et al.},
        title = "{Fast Radio Bursts as Cosmological Probes}",
    booktitle = {Advancing Astrophysics with the SKA I},
         year = 2026,
       editor = {{Bonaldi}, Anna and {Bourke}, Tyler L. and {Hartley}, Philippa and et al.},
          eid = {124},
        pages = {124},
          doi = {10.48550/arXiv.2606.27714},
       adsurl = {https://scixplorer.org/abs/2026aask.bookE.124C}
}

@ARTICLE{Wang2016,
       author = {{Wang}, Jie-Shuang and {Yang}, Yuan-Pei and {Wu}, Xue-Feng and {Dai}, Zi-Gao and {Wang}, Fa-Yin},
        title = "{Fast Radio Bursts from the Inspiral of Double Neutron Stars}",
      journal = {\apjl},
         year = 2016,
        month = may,
       volume = {822},
       number = {1},
          eid = {L7},
        pages = {L7},
          doi = {10.3847/2041-8205/822/1/L7},
archivePrefix = {arXiv},
       eprint = {1603.02014},
 primaryClass = {astro-ph.HE},
       adsurl = {https://ui.adsabs.harvard.edu/abs/2016ApJ...822L...7W}
}

@ARTICLE{petroff2022,
       author = {{Petroff}, E. and {Hessels}, J.~W.~T. and {Lorimer}, D.~R.},
        title = "{Fast radio bursts at the dawn of the 2020s}",
      journal = {\aapr},
         year = 2022,
        month = dec,
       volume = {30},
       number = {1},
          eid = {2},
        pages = {2},
          doi = {10.1007/s00159-022-00139-w},
archivePrefix = {arXiv},
       eprint = {2107.10113},
 primaryClass = {astro-ph.HE},
       adsurl = {https://ui.adsabs.harvard.edu/abs/2022A&ARv..30....2P}
}

@ARTICLE{zhang2023,
       author = {{Zhang}, Bing},
        title = "{The physics of fast radio bursts}",
      journal = {Reviews of Modern Physics},
         year = 2023,
        month = jul,
       volume = {95},
       number = {3},
          eid = {035005},
        pages = {035005},
          doi = {10.1103/RevModPhys.95.035005},
archivePrefix = {arXiv},
       eprint = {2212.03972},
 primaryClass = {astro-ph.HE},
       adsurl = {https://ui.adsabs.harvard.edu/abs/2023RvMP...95c5005Z}
}

@ARTICLE{CHIME_catalog1,
       author = {{CHIME/FRB Collaboration} and {Amiri}, Mandana and {Andersen}, Bridget C. and {Bandura}, Kevin and {Berger}, Sabrina and {Bhardwaj}, Mohit and {Boyce}, Michelle M. and {Boyle}, P.~J. and {Brar}, Charanjot and {Breitman}, Daniela and {Cassanelli}, Tomas and {Chawla}, Pragya and {Chen}, Tianyue and {Cliche}, J.-F. and {Cook}, Amanda and {Cubranic}, Davor and {Curtin}, Alice P. and {Deng}, Meiling and {Dobbs}, Matt and {Dong}, Fengqiu Adam and {Eadie}, Gwendolyn and {Fandino}, Mateus and {Fonseca}, Emmanuel and {Gaensler}, B.~M. and {Giri}, Utkarsh and {Good}, Deborah C. and {Halpern}, Mark and {Hill}, Alex S. and {Hinshaw}, Gary and {Josephy}, Alexander and {Kaczmarek}, Jane F. and {Kader}, Zarif and {Kania}, Joseph W. and {Kaspi}, Victoria M. and {Landecker}, T.~L. and {Lang}, Dustin and {Leung}, Calvin and {Li}, Dongzi and {Lin}, Hsiu-Hsien and {Masui}, Kiyoshi W. and {McKinven}, Ryan and {Mena-Parra}, Juan and {Merryfield}, Marcus and {Meyers}, Bradley W. and {Michilli}, Daniele and {Milutinovic}, Nikola and {Mirhosseini}, Arash and {M{\"u}nchmeyer}, Moritz and {Naidu}, Arun and {Newburgh}, Laura and {Ng}, Cherry and {Patel}, Chitrang and {Pen}, Ue-Li and {Petroff}, Emily and {Pinsonneault-Marotte}, Tristan and {Pleunis}, Ziggy and {Rafiei-Ravandi}, Masoud and {Rahman}, Mubdi and {Ransom}, Scott M. and {Renard}, Andre and {Sanghavi}, Pranav and {Scholz}, Paul and {Shaw}, J. Richard and {Shin}, Kaitlyn and {Siegel}, Seth R. and {Sikora}, Andrew E. and {Singh}, Saurabh and {Smith}, Kendrick M. and {Stairs}, Ingrid and {Tan}, Chia Min and {Tendulkar}, S.~P. and {Vanderlinde}, Keith and {Wang}, Haochen and {Wulf}, Dallas and {Zwaniga}, A.~V.},
        title = "{The First CHIME/FRB Fast Radio Burst Catalog}",
      journal = {\apjs},
         year = 2021,
        month = dec,
       volume = {257},
       number = {2},
          eid = {59},
        pages = {59},
          doi = {10.3847/1538-4365/ac33ab},
archivePrefix = {arXiv},
       eprint = {2106.04352},
 primaryClass = {astro-ph.HE},
       adsurl = {https://ui.adsabs.harvard.edu/abs/2021ApJS..257...59C}
}

@ARTICLE{CHIME_catalog2,
       author = {{Chime/Frb Collaboration} and {Abbott}, Thomas and {Andersen}, Bridget C. and {Andrew}, Shion and {Bandura}, Kevin and {Bhardwaj}, Mohit and {Bhusare}, Yash and {Brar}, Charanjot and {Cassanelli}, Tomas and {Chatterjee}, Shami and {Cliche}, Jean-Francois and {Cook}, Amanda M. and {Curtin}, Alice and {Dobbs}, Matt and {Dong}, Fengqiu Adam and {Eadie}, Gwendolyn and {Eftekhari}, Tarraneh and {Fonseca}, Emmanuel and {Gaensler}, B.~M. and {Good}, Deborah and {Halpern}, Mark and {Hessels}, Jason W.~T. and {Ibik}, Adaeze and {Jain}, Naman and {Joseph}, Ronniy C. and {Kader}, Zarif and {Kaspi}, Victoria M. and {Khan}, Afrokk and {Kharel}, Bikash and {Kumar}, Ajay and {Landecker}, T.~L. and {Lang}, Dustin and {Lanman}, Adam E. and {L'Argent}, Magnus and {Lazda}, Mattias and {Leung}, Calvin and {Li}, Dong Zi and {Lintott}, Chris J. and {Main}, Robert and {Masui}, Kiyoshi W. and {Mate}, Sujay and {McGregor}, Kyle and {McKinven}, Ryan and {Mena-Parra}, Juan and {Meyers}, Bradley W. and {Michilli}, Daniele and {Ng}, Cherry and {Ng}, Mason and {Nimmo}, Kenzie and {Noble}, Gavin and {Pandhi}, Ayush and {Patil}, Swarali S. and {Pearlman}, Aaron B. and {Pen}, Ue-Li and {Pleunis}, Ziggy and {Prochaska}, J. Xavier and {Rafiei-Ravandi}, Masoud and {Ransom}, Scott and {Renard}, Andre and {Sammons}, Mawson W. and {Sand}, Ketan R. and {Scholz}, Paul and {Shah}, Vishwangi and {Shin}, Kaitlyn and {Siegel}, Seth R. and {Sirota}, Sloane and {Smith}, Kendrick and {Stairs}, Ingrid and {Stenning}, David C. and {Tendulkar}, Shriharsh P. and {Vanderlinde}, Keith and {Walmsley}, Mike and {Wang}, Haochen and {Wulf}, Dallas},
        title = "{The Second CHIME/FRB Catalog of Fast Radio Bursts}",
      journal = {\apjs},
         year = 2026,
        month = mar,
       volume = {283},
       number = {1},
          eid = {34},
        pages = {34},
          doi = {10.3847/1538-4365/ae3828},
archivePrefix = {arXiv},
       eprint = {2601.09399},
 primaryClass = {astro-ph.HE},
       adsurl = {https://ui.adsabs.harvard.edu/abs/2026ApJS..283...34C}
}

@ARTICLE{Shin2023,
       author = {{Shin}, Kaitlyn and {Masui}, Kiyoshi W. and {Bhardwaj}, Mohit and {Cassanelli}, Tomas and {Chawla}, Pragya and {Dobbs}, Matt and {Dong}, Fengqiu Adam and {Fonseca}, Emmanuel and {Gaensler}, B.~M. and {Herrera-Mart{\'\i}n}, Antonio and {Kaczmarek}, Jane and {Kaspi}, Victoria and {Leung}, Calvin and {Merryfield}, Marcus and {Michilli}, Daniele and {M{\"u}nchmeyer}, Moritz and {Pearlman}, Aaron B. and {Rafiei-Ravandi}, Masoud and {Smith}, Kendrick and {Stairs}, Ingrid and {Tendulkar}, Shriharsh P.},
        title = "{Inferring the Energy and Distance Distributions of Fast Radio Bursts Using the First CHIME/FRB Catalog}",
      journal = {\apj},
         year = 2023,
        month = feb,
       volume = {944},
       number = {1},
          eid = {105},
        pages = {105},
          doi = {10.3847/1538-4357/acaf06},
archivePrefix = {arXiv},
       eprint = {2207.14316},
 primaryClass = {astro-ph.HE},
       adsurl = {https://ui.adsabs.harvard.edu/abs/2023ApJ...944..105S}
}

@ARTICLE{2026PhRvL.137d1001W,
       author = {{Wang}, Haochen and {Masui}, Kiyoshi and {Andrew}, Shion and {Bhardwaj}, Mohit and {Fonseca}, Emmanuel and {Gaensler}, B.~M. and {Joseph}, R.~C. and {Kaspi}, Victoria M. and {Kharel}, Bikash and {Lanman}, Adam E. and et al.},
        title = "{Measurement of the Dispersion-Galaxy Cross-Power Spectrum with the Second CHIME/FRB Catalog}",
      journal = {\prl},
         year = 2026,
        month = jul,
       volume = {137},
       number = {4},
          eid = {041001},
        pages = {041001},
          doi = {10.1103/9th9-qc51},
archivePrefix = {arXiv},
       eprint = {2506.08932},
 primaryClass = {astro-ph.CO},
       adsurl = {https://ui.adsabs.harvard.edu/abs/2026PhRvL.137d1001W}
}

@ARTICLE{2020Natur.587...54C,
       author = {{CHIME/FRB Collaboration} and {Andersen}, B.~C. and {Bandura}, K.~M. and {Bhardwaj}, M. and {Bij}, A. and {Boyce}, M.~M. and {Boyle}, P.~J. and {Brar}, C. and {Cassanelli}, T. and {Chawla}, P. and {Chen}, T. and {Cliche}, J.-F. and {Cook}, A. and {Cubranic}, D. and {Curtin}, A.~P. and {Denman}, N.~T. and {Dobbs}, M. and {Dong}, F.~Q. and {Fandino}, M. and {Fonseca}, E. and {Gaensler}, B.~M. and {Giri}, U. and {Good}, D.~C. and {Halpern}, M. and {Hill}, A.~S. and {Hinshaw}, G.~F. and {H{\"o}fer}, C. and {Josephy}, A. and {Kania}, J.~W. and {Kaspi}, V.~M. and {Landecker}, T.~L. and {Leung}, C. and {Li}, D.~Z. and {Lin}, H.-H. and {Masui}, K.~W. and {McKinven}, R. and {Mena-Parra}, J. and {Merryfield}, M. and {Meyers}, B.~W. and {Michilli}, D. and {Milutinovic}, N. and {Mirhosseini}, A. and {M{\"u}nchmeyer}, M. and {Naidu}, A. and {Newburgh}, L.~B. and {Ng}, C. and {Patel}, C. and {Pen}, U.-L. and {Pinsonneault-Marotte}, T. and {Pleunis}, Z. and {Quine}, B.~M. and {Rafiei-Ravandi}, M. and {Rahman}, M. and {Ransom}, S.~M. and {Renard}, A. and {Sanghavi}, P. and {Scholz}, P. and {Shaw}, J.~R. and {Shin}, K. and {Siegel}, S.~R. and {Singh}, S. and {Smegal}, R.~J. and {Smith}, K.~M. and {Stairs}, I.~H. and {Tan}, C.~M. and {Tendulkar}, S.~P. and {Tretyakov}, I. and {Vanderlinde}, K. and {Wang}, H. and {Wulf}, D. and {Zwaniga}, A.~V.},
        title = "{A bright millisecond-duration radio burst from a Galactic magnetar}",
      journal = {\nat},
         year = 2020,
        month = nov,
       volume = {587},
       number = {7832},
        pages = {54-58},
          doi = {10.1038/s41586-020-2863-y},
archivePrefix = {arXiv},
       eprint = {2005.10324},
 primaryClass = {astro-ph.HE},
       adsurl = {https://ui.adsabs.harvard.edu/abs/2020Natur.587...54C}
}

@ARTICLE{2020Natur.587...59B,
       author = {{Bochenek}, C.~D. and {Ravi}, V. and {Belov}, K.~V. and {Hallinan}, G. and {Kocz}, J. and {Kulkarni}, S.~R. and {McKenna}, D.~L.},
        title = "{A fast radio burst associated with a Galactic magnetar}",
      journal = {\nat},
         year = 2020,
        month = nov,
       volume = {587},
       number = {7832},
        pages = {59-62},
          doi = {10.1038/s41586-020-2872-x},
archivePrefix = {arXiv},
       eprint = {2005.10828},
 primaryClass = {astro-ph.HE},
       adsurl = {https://ui.adsabs.harvard.edu/abs/2020Natur.587...59B}
}

@ARTICLE{2021ApJ...910L..18B,
       author = {{Bhardwaj}, M. and {Gaensler}, B.~M. and {Kaspi}, V.~M. and {Landecker}, T.~L. and {Mckinven}, R. and {Michilli}, D. and {Pleunis}, Z. and {Tendulkar}, S.~P. and {Andersen}, B.~C. and {Boyle}, P.~J. and {Cassanelli}, T. and {Chawla}, P. and {Cook}, A. and {Dobbs}, M. and {Fonseca}, E. and {Kaczmarek}, J. and {Leung}, C. and {Masui}, K. and {Mnchmeyer}, M. and {Ng}, C. and {Rafiei-Ravandi}, M. and {Scholz}, P. and {Shin}, K. and {Smith}, K.~M. and {Stairs}, I.~H. and {Zwaniga}, A.~V.},
        title = "{A Nearby Repeating Fast Radio Burst in the Direction of M81}",
      journal = {\apjl},
         year = 2021,
        month = apr,
       volume = {910},
       number = {2},
          eid = {L18},
        pages = {L18},
          doi = {10.3847/2041-8213/abeaa6},
archivePrefix = {arXiv},
       eprint = {2103.01295},
 primaryClass = {astro-ph.HE},
       adsurl = {https://ui.adsabs.harvard.edu/abs/2021ApJ...910L..18B}
}

@ARTICLE{2022Natur.602..585K,
       author = {{Kirsten}, F. and {Marcote}, B. and {Nimmo}, K. and {Hessels}, J.~W.~T. and {Bhardwaj}, M. and {Tendulkar}, S.~P. and {Keimpema}, A. and {Yang}, J. and {Snelders}, M.~P. and {Scholz}, P. and {Pearlman}, A.~B. and {Law}, C.~J. and {Peters}, W.~M. and {Giroletti}, M. and {Paragi}, Z. and {Bassa}, C. and {Hewitt}, D.~M. and {Bach}, U. and {Bezrukovs}, V. and {Burgay}, M. and {Buttaccio}, S.~T. and {Conway}, J.~E. and {Corongiu}, A. and {Feiler}, R. and {Forss{\'e}n}, O. and {Gawro{\'n}ski}, M.~P. and {Karuppusamy}, R. and {Kharinov}, M.~A. and {Lindqvist}, M. and {Maccaferri}, G. and {Melnikov}, A. and {Ould-Boukattine}, O.~S. and {Possenti}, A. and {Surcis}, G. and {Wang}, N. and {Yuan}, J. and {Aggarwal}, K. and {Anna-Thomas}, R. and {Bower}, G.~C. and {Blaauw}, R. and {Burke-Spolaor}, S. and {Cassanelli}, T. and {Clarke}, T.~E. and {Fonseca}, E. and {Gaensler}, B.~M. and {Gopinath}, A. and {Kaspi}, V.~M. and {Kassim}, N. and {Lazio}, T.~J.~W. and {Leung}, C. and {Li}, D.~Z. and {Lin}, H.~H. and {Masui}, K.~W. and {Mckinven}, R. and {Michilli}, D. and {Mikhailov}, A.~G. and {Ng}, C. and {Orbidans}, A. and {Pen}, U.~L. and {Petroff}, E. and {Rahman}, M. and {Ransom}, S.~M. and {Shin}, K. and {Smith}, K.~M. and {Stairs}, I.~H. and {Vlemmings}, W.},
        title = "{A repeating fast radio burst source in a globular cluster}",
      journal = {\nat},
         year = 2022,
        month = feb,
       volume = {602},
       number = {7898},
        pages = {585-589},
          doi = {10.1038/s41586-021-04354-w},
archivePrefix = {arXiv},
       eprint = {2105.11445},
 primaryClass = {astro-ph.HE},
       adsurl = {https://ui.adsabs.harvard.edu/abs/2022Natur.602..585K}
}

@ARTICLE{2018ApJ...863...48C,
       author = {{CHIME/FRB Collaboration} and {Amiri}, M. and {Bandura}, K. and {Berger}, P. and {Bhardwaj}, M. and {Boyce}, M.~M. and {Boyle}, P.~J. and {Brar}, C. and {Burhanpurkar}, M. and {Chawla}, P. and {Chowdhury}, J. and {Cliche}, J.-F. and {Cranmer}, M.~D. and {Cubranic}, D. and {Deng}, M. and {Denman}, N. and {Dobbs}, M. and {Fandino}, M. and {Fonseca}, E. and {Gaensler}, B.~M. and {Giri}, U. and {Gilbert}, A.~J. and {Good}, D.~C. and {Guliani}, S. and {Halpern}, M. and {Hinshaw}, G. and {H{\"o}fer}, C. and {Josephy}, A. and {Kaspi}, V.~M. and {Landecker}, T.~L. and {Lang}, D. and {Liao}, H. and {Masui}, K.~W. and {Mena-Parra}, J. and {Naidu}, A. and {Newburgh}, L.~B. and {Ng}, C. and {Patel}, C. and {Pen}, U.-L. and {Pinsonneault-Marotte}, T. and {Pleunis}, Z. and {Rafiei Ravandi}, M. and {Ransom}, S.~M. and {Renard}, A. and {Scholz}, P. and {Sigurdson}, K. and {Siegel}, S.~R. and {Smith}, K.~M. and {Stairs}, I.~H. and {Tendulkar}, S.~P. and {Vanderlinde}, K. and {Wiebe}, D.~V.},
        title = "{The CHIME Fast Radio Burst Project: System Overview}",
      journal = {\apj},
         year = 2018,
        month = aug,
       volume = {863},
       number = {1},
          eid = {48},
        pages = {48},
          doi = {10.3847/1538-4357/aad188},
archivePrefix = {arXiv},
       eprint = {1803.11235},
 primaryClass = {astro-ph.IM},
       adsurl = {https://ui.adsabs.harvard.edu/abs/2018ApJ...863...48C}
}

@ARTICLE{2024A&A...690A.377W,
       author = {{Wang}, Yuyang and {van Leeuwen}, Joeri},
        title = "{Birth and evolution of fast radio bursts: Strong population-based evidence for a neutron-star origin}",
      journal = {\aap},
         year = 2024,
        month = oct,
       volume = {690},
          eid = {A377},
        pages = {A377},
          doi = {10.1051/0004-6361/202450673},
archivePrefix = {arXiv},
       eprint = {2405.06281},
 primaryClass = {astro-ph.HE},
       adsurl = {https://ui.adsabs.harvard.edu/abs/2024A&A...690A.377W}
}

@ARTICLE{2022ApJ...924L..14Z,
       author = {{Zhang}, Rachel C. and {Zhang}, Bing},
        title = "{The CHIME Fast Radio Burst Population Does Not Track the Star Formation History of the Universe}",
      journal = {\apjl},
         year = 2022,
        month = jan,
       volume = {924},
       number = {1},
          eid = {L14},
        pages = {L14},
          doi = {10.3847/2041-8213/ac46ad},
archivePrefix = {arXiv},
       eprint = {2109.07558},
 primaryClass = {astro-ph.HE},
       adsurl = {https://ui.adsabs.harvard.edu/abs/2022ApJ...924L..14Z}
}

@ARTICLE{2019MNRAS.487.3672Z,
       author = {{Zhang}, G.~Q. and {Wang}, F.~Y.},
        title = "{Energy function, formation rate, and low-metallicity environment of fast radio bursts}",
      journal = {\mnras},
         year = 2019,
        month = aug,
       volume = {487},
       number = {3},
        pages = {3672-3678},
          doi = {10.1093/mnras/stz1566},
archivePrefix = {arXiv},
       eprint = {1906.01176},
 primaryClass = {astro-ph.HE},
       adsurl = {https://ui.adsabs.harvard.edu/abs/2019MNRAS.487.3672Z}
}

@ARTICLE{Chenjh2024,
       author = {{Chen}, J.~H. and {Jia}, X.~D. and {Dong}, X.~F. and {Wang}, F.~Y.},
        title = "{The Formation Rate and Luminosity Function of Fast Radio Bursts}",
      journal = {\apjl},
         year = 2024,
        month = oct,
       volume = {973},
       number = {2},
          eid = {L54},
        pages = {L54},
          doi = {10.3847/2041-8213/ad7b39},
archivePrefix = {arXiv},
       eprint = {2406.03672},
 primaryClass = {astro-ph.HE},
       adsurl = {https://ui.adsabs.harvard.edu/abs/2024ApJ...973L..54C}
}

@ARTICLE{Jia2026,
       author = {{Jia}, X.~D. and {Gao}, D.~H. and {Chen}, J.~H. and {Wu}, Q. and {Yi}, S.~X. and {Wang}, F.~Y.},
        title = "{Cosmological Evolution of Fast Radio Bursts and Its Rapid Decline Relative to Star Formation Rate}",
      journal = {\apj},
         year = 2026,
        month = jun,
       volume = {1003},
       number = {2},
          eid = {179},
        pages = {179},
          doi = {10.3847/1538-4357/ae665c},
archivePrefix = {arXiv},
       eprint = {2604.26574},
 primaryClass = {astro-ph.HE},
       adsurl = {https://ui.adsabs.harvard.edu/abs/2026ApJ..1003..179J}
}

@ARTICLE{2026arXiv260704792Z,
       author = {{Zhang}, Zi-Liang and {Zhang}, Bing},
        title = "{Evidence for a Delayed Progenitor Population for CHIME non-repeating Fast Radio Bursts from a Self-Consistent Forward and Backward Inference Framework}",
      journal = {arXiv e-prints},
         year = 2026,
        month = jul,
          eid = {arXiv:2607.04792},
        pages = {arXiv:2607.04792},
          doi = {10.48550/arXiv.2607.04792},
archivePrefix = {arXiv},
       eprint = {2607.04792},
 primaryClass = {astro-ph.HE},
       adsurl = {https://ui.adsabs.harvard.edu/abs/2026arXiv260704792Z}
}

@ARTICLE{2022MNRAS.511.1961H,
       author = {{Hashimoto}, Tetsuya and {Goto}, Tomotsugu and {Chen}, Bo Han and {Ho}, Simon C.-C. and {Hsiao}, Tiger Y.-Y. and {Wong}, Yi Hang Valerie and {On}, Alvina Y.~L. and {Kim}, Seong Jin and {Kilerci-Eser}, Ece and {Huang}, Kai-Chun and {Santos}, Daryl Joe D. and {Yamasaki}, Shotaro},
        title = "{Energy functions of fast radio bursts derived from the first CHIME/FRB catalogue}",
      journal = {\mnras},
         year = 2022,
        month = apr,
       volume = {511},
       number = {2},
        pages = {1961-1976},
          doi = {10.1093/mnras/stac065},
archivePrefix = {arXiv},
       eprint = {2201.03574},
 primaryClass = {astro-ph.HE},
       adsurl = {https://ui.adsabs.harvard.edu/abs/2022MNRAS.511.1961H}
}

@ARTICLE{2022JCAP...01..040Q,
       author = {{Qiang}, Da-Chun and {Li}, Shu-Ling and {Wei}, Hao},
        title = "{Fast radio burst distributions consistent with the first CHIME/FRB catalog}",
      journal = {\jcap},
         year = 2022,
        month = jan,
       volume = {2022},
       number = {1},
          eid = {040},
        pages = {040},
          doi = {10.1088/1475-7516/2022/01/040},
archivePrefix = {arXiv},
       eprint = {2111.07476},
 primaryClass = {astro-ph.HE},
       adsurl = {https://ui.adsabs.harvard.edu/abs/2022JCAP...01..040Q}
}

@ARTICLE{2025ApJ...988L..64C,
       author = {{Champati}, Sujay and {Petrosian}, Vah{\'e}},
        title = "{Cosmological Evolution of Fast Radio Bursts and the Star Formation Rate}",
      journal = {\apjl},
         year = 2025,
        month = aug,
       volume = {988},
       number = {2},
          eid = {L64},
        pages = {L64},
          doi = {10.3847/2041-8213/adefdb},
archivePrefix = {arXiv},
       eprint = {2504.13343},
 primaryClass = {astro-ph.CO},
       adsurl = {https://ui.adsabs.harvard.edu/abs/2025ApJ...988L..64C}
}

@ARTICLE{2025ApJ...986..100G,
       author = {{Gupta}, Om and {Beniamini}, Paz and {Kumar}, Pawan and {Finkelstein}, Steven L.},
        title = "{The Cosmic Evolution of Fast Radio Bursts Inferred from the CHIME/FRB Baseband Catalog 1}",
      journal = {\apj},
         year = 2025,
        month = jun,
       volume = {986},
       number = {1},
          eid = {100},
        pages = {100},
          doi = {10.3847/1538-4357/add14c},
archivePrefix = {arXiv},
       eprint = {2501.09810},
 primaryClass = {astro-ph.HE},
       adsurl = {https://ui.adsabs.harvard.edu/abs/2025ApJ...986..100G}
}

@ARTICLE{2022NatCo..13.4382W,
       author = {{Wang}, F.~Y. and {Zhang}, G.~Q. and {Dai}, Z.~G. and {Cheng}, K.~S.},
        title = "{Repeating fast radio burst 20201124A originates from a magnetar/Be star binary}",
      journal = {Nature Communications},
         year = 2022,
        month = sep,
       volume = {13},
          eid = {4382},
        pages = {4382},
          doi = {10.1038/s41467-022-31923-y},
archivePrefix = {arXiv},
       eprint = {2204.08124},
 primaryClass = {astro-ph.HE},
       adsurl = {https://ui.adsabs.harvard.edu/abs/2022NatCo..13.4382W}
}

@ARTICLE{2026ApJ..1009..211W,
       author = {{Wang}, F.~Y. and {Lan}, H.~T. and {Zhao}, Z.~Y. and {Wu}, Q. and {Feng}, Y. and {Yi}, S.~X. and {Dai}, Z.~G. and {Cheng}, K.~S.},
        title = "{A Unified Model of Active Repeating Fast Radio Bursts Associated with Persistent Radio Sources}",
      journal = {\apj},
         year = 2026,
        month = oct,
       volume = {1009},
       number = {2},
          eid = {211},
        pages = {211},
          doi = {10.3847/1538-4357/ae9d5a},
       adsurl = {https://ui.adsabs.harvard.edu/abs/2026ApJ..1009..211W}
}

@ARTICLE{2018ApJ...858...89C,
       author = {{Cao}, Xiao-Feng and {Yu}, Yun-Wei and {Zhou}, Xia},
        title = "{Compact Binary Mergers and the Event Rate of Fast Radio Bursts}",
      journal = {\apj},
         year = 2018,
        month = may,
       volume = {858},
       number = {2},
          eid = {89},
        pages = {89},
          doi = {10.3847/1538-4357/aabadd},
archivePrefix = {arXiv},
       eprint = {1803.06266},
 primaryClass = {astro-ph.HE},
       adsurl = {https://ui.adsabs.harvard.edu/abs/2018ApJ...858...89C}
}

@ARTICLE{2014A&A...562A.137F,
       author = {{Falcke}, Heino and {Rezzolla}, Luciano},
        title = "{Fast radio bursts: the last sign of supramassive neutron stars}",
      journal = {\aap},
         year = 2014,
        month = feb,
       volume = {562},
          eid = {A137},
        pages = {A137},
          doi = {10.1051/0004-6361/201321996},
archivePrefix = {arXiv},
       eprint = {1307.1409},
 primaryClass = {astro-ph.HE},
       adsurl = {https://ui.adsabs.harvard.edu/abs/2014A&A...562A.137F}
}

@ARTICLE{2014MNRAS.441.2433R,
       author = {{Ravi}, Vikram and {Lasky}, Paul D.},
        title = "{The birth of black holes: neutron star collapse times, gamma-ray bursts and fast radio bursts}",
      journal = {\mnras},
         year = 2014,
        month = jul,
       volume = {441},
       number = {3},
        pages = {2433-2439},
          doi = {10.1093/mnras/stu720},
archivePrefix = {arXiv},
       eprint = {1403.6327},
 primaryClass = {astro-ph.HE},
       adsurl = {https://ui.adsabs.harvard.edu/abs/2014MNRAS.441.2433R}
}

@ARTICLE{2019ApJ...886..110M,
       author = {{Margalit}, Ben and {Berger}, Edo and {Metzger}, Brian D.},
        title = "{Fast Radio Bursts from Magnetars Born in Binary Neutron Star Mergers and Accretion Induced Collapse}",
      journal = {\apj},
         year = 2019,
        month = dec,
       volume = {886},
       number = {2},
          eid = {110},
        pages = {110},
          doi = {10.3847/1538-4357/ab4c31},
archivePrefix = {arXiv},
       eprint = {1907.00016},
 primaryClass = {astro-ph.HE},
       adsurl = {https://ui.adsabs.harvard.edu/abs/2019ApJ...886..110M}
}

@ARTICLE{2017ApJ...835...29Y,
       author = {{Yao}, J.~M. and {Manchester}, R.~N. and {Wang}, N.},
        title = "{A New Electron-density Model for Estimation of Pulsar and FRB Distances}",
      journal = {\apj},
         year = 2017,
        month = jan,
       volume = {835},
       number = {1},
          eid = {29},
        pages = {29},
          doi = {10.3847/1538-4357/835/1/29},
archivePrefix = {arXiv},
       eprint = {1610.09448},
 primaryClass = {astro-ph.GA},
       adsurl = {https://ui.adsabs.harvard.edu/abs/2017ApJ...835...29Y}
}

@ARTICLE{2002astro.ph..7156C,
       author = {{Cordes}, J.~M. and {Lazio}, T.~J.~W.},
        title = "{NE2001.I. A New Model for the Galactic Distribution of Free Electrons and its Fluctuations}",
      journal = {arXiv e-prints},
         year = 2002,
        month = jul,
          eid = {astro-ph/0207156},
        pages = {astro-ph/0207156},
          doi = {10.48550/arXiv.astro-ph/0207156},
archivePrefix = {arXiv},
       eprint = {astro-ph/0207156},
 primaryClass = {astro-ph},
       adsurl = {https://ui.adsabs.harvard.edu/abs/2002astro.ph..7156C}
}

@ARTICLE{2015MNRAS.451.4277D,
       author = {{Dolag}, K. and {Gaensler}, B.~M. and {Beck}, A.~M. and {Beck}, M.~C.},
        title = "{Constraints on the distribution and energetics of fast radio bursts using cosmological hydrodynamic simulations}",
      journal = {\mnras},
         year = 2015,
        month = aug,
       volume = {451},
       number = {4},
        pages = {4277-4289},
          doi = {10.1093/mnras/stv1190},
archivePrefix = {arXiv},
       eprint = {1412.4829},
 primaryClass = {astro-ph.CO},
       adsurl = {https://ui.adsabs.harvard.edu/abs/2015MNRAS.451.4277D}
}

@ARTICLE{2009RvMP...81.1405M,
       author = {{Meiksin}, Avery A.},
        title = "{The physics of the intergalactic medium}",
      journal = {Reviews of Modern Physics},
         year = 2009,
        month = oct,
       volume = {81},
       number = {4},
        pages = {1405-1469},
          doi = {10.1103/RevModPhys.81.1405},
archivePrefix = {arXiv},
       eprint = {0711.3358},
 primaryClass = {astro-ph},
       adsurl = {https://ui.adsabs.harvard.edu/abs/2009RvMP...81.1405M}
}

@ARTICLE{2011MNRAS.410.1096B,
       author = {{Becker}, George D. and {Bolton}, James S. and {Haehnelt}, Martin G. and {Sargent}, Wallace L.~W.},
        title = "{Detection of extended He II reionization in the temperature evolution of the intergalactic medium}",
      journal = {\mnras},
         year = 2011,
        month = jan,
       volume = {410},
       number = {2},
        pages = {1096-1112},
          doi = {10.1111/j.1365-2966.2010.17507.x},
archivePrefix = {arXiv},
       eprint = {1008.2622},
 primaryClass = {astro-ph.CO},
       adsurl = {https://ui.adsabs.harvard.edu/abs/2011MNRAS.410.1096B}
}

@ARTICLE{2020A&A...641A...6P,
       author = {{Planck Collaboration} and {Aghanim}, N. and {Akrami}, Y. and {Ashdown}, M. and {Aumont}, J. and {Baccigalupi}, C. and {Ballardini}, M. and {Banday}, A.~J. and {Barreiro}, R.~B. and {Bartolo}, N. and {Basak}, S. and {Battye}, R. and {Benabed}, K. and {Bernard}, J.-P. and {Bersanelli}, M. and {Bielewicz}, P. and {Bock}, J.~J. and {Bond}, J.~R. and {Borrill}, J. and {Bouchet}, F.~R. and {Boulanger}, F. and {Bucher}, M. and {Burigana}, C. and {Butler}, R.~C. and {Calabrese}, E. and {Cardoso}, J.-F. and {Carron}, J. and {Challinor}, A. and {Chiang}, H.~C. and {Chluba}, J. and {Colombo}, L.~P.~L. and {Combet}, C. and {Contreras}, D. and {Crill}, B.~P. and {Cuttaia}, F. and {de Bernardis}, P. and {de Zotti}, G. and {Delabrouille}, J. and {Delouis}, J.-M. and {Di Valentino}, E. and {Diego}, J.~M. and {Dor{\'e}}, O. and {Douspis}, M. and {Ducout}, A. and {Dupac}, X. and {Dusini}, S. and {Efstathiou}, G. and {Elsner}, F. and {En{\ss}lin}, T.~A. and {Eriksen}, H.~K. and {Fantaye}, Y. and {Farhang}, M. and {Fergusson}, J. and {Fernandez-Cobos}, R. and {Finelli}, F. and {Forastieri}, F. and {Frailis}, M. and {Fraisse}, A.~A. and {Franceschi}, E. and {Frolov}, A. and {Galeotta}, S. and {Galli}, S. and {Ganga}, K. and {G{\'e}nova-Santos}, R.~T. and {Gerbino}, M. and {Ghosh}, T. and {Gonz{\'a}lez-Nuevo}, J. and {G{\'o}rski}, K.~M. and {Gratton}, S. and {Gruppuso}, A. and {Gudmundsson}, J.~E. and {Hamann}, J. and {Handley}, W. and {Hansen}, F.~K. and {Herranz}, D. and {Hildebrandt}, S.~R. and {Hivon}, E. and {Huang}, Z. and {Jaffe}, A.~H. and {Jones}, W.~C. and {Karakci}, A. and {Keih{\"a}nen}, E. and {Keskitalo}, R. and {Kiiveri}, K. and {Kim}, J. and {Kisner}, T.~S. and {Knox}, L. and {Krachmalnicoff}, N. and {Kunz}, M. and {Kurki-Suonio}, H. and {Lagache}, G. and {Lamarre}, J.-M. and {Lasenby}, A. and {Lattanzi}, M. and {Lawrence}, C.~R. and {Le Jeune}, M. and {Lemos}, P. and {Lesgourgues}, J. and {Levrier}, F. and {Lewis}, A. and {Liguori}, M. and {Lilje}, P.~B. and {Lilley}, M. and {Lindholm}, V. and {L{\'o}pez-Caniego}, M. and {Lubin}, P.~M. and {Ma}, Y.-Z. and {Mac{\'\i}as-P{\'e}rez}, J.~F. and {Maggio}, G. and {Maino}, D. and {Mandolesi}, N. and {Mangilli}, A. and {Marcos-Caballero}, A. and {Maris}, M. and {Martin}, P.~G. and {Martinelli}, M. and {Mart{\'\i}nez-Gonz{\'a}lez}, E. and {Matarrese}, S. and {Mauri}, N. and {McEwen}, J.~D. and {Meinhold}, P.~R. and {Melchiorri}, A. and {Mennella}, A. and {Migliaccio}, M. and {Millea}, M. and {Mitra}, S. and {Miville-Desch{\^e}nes}, M.-A. and {Molinari}, D. and {Montier}, L. and {Morgante}, G. and {Moss}, A. and {Natoli}, P. and {N{\o}rgaard-Nielsen}, H.~U. and {Pagano}, L. and {Paoletti}, D. and {Partridge}, B. and {Patanchon}, G. and {Peiris}, H.~V. and {Perrotta}, F. and {Pettorino}, V. and {Piacentini}, F. and {Polastri}, L. and {Polenta}, G. and {Puget}, J.-L. and {Rachen}, J.~P. and {Reinecke}, M. and {Remazeilles}, M. and {Renzi}, A. and {Rocha}, G. and {Rosset}, C. and {Roudier}, G. and {Rubi{\~n}o-Mart{\'\i}n}, J.~A. and {Ruiz-Granados}, B. and {Salvati}, L. and {Sandri}, M. and {Savelainen}, M. and {Scott}, D. and {Shellard}, E.~P.~S. and {Sirignano}, C. and {Sirri}, G. and {Spencer}, L.~D. and {Sunyaev}, R. and {Suur-Uski}, A.-S. and {Tauber}, J.~A. and {Tavagnacco}, D. and {Tenti}, M. and {Toffolatti}, L. and {Tomasi}, M. and {Trombetti}, T. and {Valenziano}, L. and {Valiviita}, J. and {Van Tent}, B. and {Vibert}, L. and {Vielva}, P. and {Villa}, F. and {Vittorio}, N. and {Wandelt}, B.~D. and {Wehus}, I.~K. and {White}, M. and {White}, S.~D.~M. and {Zacchei}, A. and {Zonca}, A.},
        title = "{Planck 2018 results. VI. Cosmological parameters}",
      journal = {\aap},
         year = 2020,
        month = sep,
       volume = {641},
          eid = {A6},
        pages = {A6},
          doi = {10.1051/0004-6361/201833910},
archivePrefix = {arXiv},
       eprint = {1807.06209},
 primaryClass = {astro-ph.CO},
       adsurl = {https://ui.adsabs.harvard.edu/abs/2020A&A...641A...6P}
}

@ARTICLE{2012ApJ...759...23S,
       author = {{Shull}, J. Michael and {Smith}, Britton D. and {Danforth}, Charles W.},
        title = "{The Baryon Census in a Multiphase Intergalactic Medium: 30\% of the Baryons May Still be Missing}",
      journal = {\apj},
         year = 2012,
        month = nov,
       volume = {759},
       number = {1},
          eid = {23},
        pages = {23},
          doi = {10.1088/0004-637X/759/1/23},
archivePrefix = {arXiv},
       eprint = {1112.2706},
 primaryClass = {astro-ph.CO},
       adsurl = {https://ui.adsabs.harvard.edu/abs/2012ApJ...759...23S}
}

@ARTICLE{Yang2022,
       author = {{Yang}, K.~B. and {Wu}, Q. and {Wang}, F.~Y.},
        title = "{Finding the Missing Baryons in the Intergalactic Medium with Localized Fast Radio Bursts}",
      journal = {\apjl},
         year = 2022,
        month = dec,
       volume = {940},
       number = {2},
          eid = {L29},
        pages = {L29},
          doi = {10.3847/2041-8213/aca145},
archivePrefix = {arXiv},
       eprint = {2211.04058},
 primaryClass = {astro-ph.HE},
       adsurl = {https://ui.adsabs.harvard.edu/abs/2022ApJ...940L..29Y}
}

@ARTICLE{2020Natur.581..391M,
       author = {{Macquart}, J.-P. and {Prochaska}, J.~X. and {McQuinn}, M. and {Bannister}, K.~W. and {Bhandari}, S. and {Day}, C.~K. and {Deller}, A.~T. and {Ekers}, R.~D. and {James}, C.~W. and {Marnoch}, L. and {Os{\l}owski}, S. and {Phillips}, C. and {Ryder}, S.~D. and {Scott}, D.~R. and {Shannon}, R.~M. and {Tejos}, N.},
        title = "{A census of baryons in the Universe from localized fast radio bursts}",
      journal = {\nat},
         year = 2020,
        month = may,
       volume = {581},
       number = {7809},
        pages = {391-395},
          doi = {10.1038/s41586-020-2300-2},
archivePrefix = {arXiv},
       eprint = {2005.13161},
 primaryClass = {astro-ph.CO},
       adsurl = {https://ui.adsabs.harvard.edu/abs/2020Natur.581..391M}
}

@ARTICLE{2021ApJ...906...49Z,
       author = {{Zhang}, Z.~J. and {Yan}, K. and {Li}, C.~M. and {Zhang}, G.~Q. and {Wang}, F.~Y.},
        title = "{Intergalactic Medium Dispersion Measures of Fast Radio Bursts Estimated from IllustrisTNG Simulation and Their Cosmological Applications}",
      journal = {\apj},
         year = 2021,
        month = jan,
       volume = {906},
       number = {1},
          eid = {49},
        pages = {49},
          doi = {10.3847/1538-4357/abceb9},
archivePrefix = {arXiv},
       eprint = {2011.14494},
 primaryClass = {astro-ph.CO},
       adsurl = {https://ui.adsabs.harvard.edu/abs/2021ApJ...906...49Z}
}

@ARTICLE{2020ApJ...900..170Z,
       author = {{Zhang}, G.~Q. and {Yu}, Hai and {He}, J.~H. and {Wang}, F.~Y.},
        title = "{Dispersion Measures of Fast Radio Burst Host Galaxies Derived from IllustrisTNG Simulation}",
      journal = {\apj},
         year = 2020,
        month = sep,
       volume = {900},
       number = {2},
          eid = {170},
        pages = {170},
          doi = {10.3847/1538-4357/abaa4a},
archivePrefix = {arXiv},
       eprint = {2007.13935},
 primaryClass = {astro-ph.HE},
       adsurl = {https://ui.adsabs.harvard.edu/abs/2020ApJ...900..170Z}
}

@ARTICLE{2022MNRAS.509.4775J,
       author = {{James}, C.~W. and {Prochaska}, J.~X. and {Macquart}, J.-P. and {North-Hickey}, F.~O. and {Bannister}, K.~W. and {Dunning}, A.},
        title = "{The z-DM distribution of fast radio bursts}",
      journal = {\mnras},
         year = 2022,
        month = feb,
       volume = {509},
       number = {4},
        pages = {4775-4802},
          doi = {10.1093/mnras/stab3051},
archivePrefix = {arXiv},
       eprint = {2101.08005},
 primaryClass = {astro-ph.HE},
       adsurl = {https://ui.adsabs.harvard.edu/abs/2022MNRAS.509.4775J}
}

@ARTICLE{2014ARA&A..52..415M,
       author = {{Madau}, Piero and {Dickinson}, Mark},
        title = "{Cosmic Star-Formation History}",
      journal = {\araa},
         year = 2014,
        month = aug,
       volume = {52},
        pages = {415-486},
          doi = {10.1146/annurev-astro-081811-125615},
archivePrefix = {arXiv},
       eprint = {1403.0007},
 primaryClass = {astro-ph.CO},
       adsurl = {https://ui.adsabs.harvard.edu/abs/2014ARA&A..52..415M}
}

@ARTICLE{2013PASP..125..306F,
       author = {{Foreman-Mackey}, Daniel and {Hogg}, David W. and {Lang}, Dustin and {Goodman}, Jonathan},
        title = "{emcee: The MCMC Hammer}",
      journal = {\pasp},
         year = 2013,
        month = mar,
       volume = {125},
       number = {925},
        pages = {306},
          doi = {10.1086/670067},
archivePrefix = {arXiv},
       eprint = {1202.3665},
 primaryClass = {astro-ph.IM},
       adsurl = {https://ui.adsabs.harvard.edu/abs/2013PASP..125..306F}
}

@ARTICLE{2015MNRAS.448.3026W,
       author = {{Wanderman}, David and {Piran}, Tsvi},
        title = "{The rate, luminosity function and time delay of non-Collapsar short GRBs}",
      journal = {\mnras},
         year = 2015,
        month = apr,
       volume = {448},
       number = {4},
        pages = {3026-3037},
          doi = {10.1093/mnras/stv123},
archivePrefix = {arXiv},
       eprint = {1405.5878},
 primaryClass = {astro-ph.HE},
       adsurl = {https://ui.adsabs.harvard.edu/abs/2015MNRAS.448.3026W}
}

@ARTICLE{1976ApJ...203..297S,
       author = {{Schechter}, P.},
        title = "{An analytic expression for the luminosity function for galaxies.}",
      journal = {\apj},
         year = 1976,
        month = jan,
       volume = {203},
        pages = {297-306},
          doi = {10.1086/154079},
       adsurl = {https://ui.adsabs.harvard.edu/abs/1976ApJ...203..297S}
}

@ARTICLE{2021ApJ...909L...8N,
       author = {{Niu}, Chen-Hui and {Li}, Di and {Luo}, Rui and {Wang}, Wei-Yang and {Yao}, Jumei and {Zhang}, Bing and {Zhu}, Wei-Wei and {Wang}, Pei and {Ye}, Haoyang and {Zhang}, Yong-Kun and {Niu}, Jia-rui and {Tang}, Ning-yu and {Duan}, Ran and {Krco}, Marko and {Dai}, Shi and {Feng}, Yi and {Miao}, Chenchen and {Pan}, Zhichen and {Qian}, Lei and {Xue}, Mengyao and {Yuan}, Mao and {Yue}, Youling and {Zhang}, Lei and {Zhang}, Xinxin},
        title = "{CRAFTS for Fast Radio Bursts: Extending the Dispersion-Fluence Relation with New FRBs Detected by FAST}",
      journal = {\apjl},
         year = 2021,
        month = mar,
       volume = {909},
       number = {1},
          eid = {L8},
        pages = {L8},
          doi = {10.3847/2041-8213/abe7f0},
archivePrefix = {arXiv},
       eprint = {2102.10546},
 primaryClass = {astro-ph.HE},
       adsurl = {https://ui.adsabs.harvard.edu/abs/2021ApJ...909L...8N}
}

@ARTICLE{2018MNRAS.481.2320L,
       author = {{Luo}, Rui and {Lee}, Kejia and {Lorimer}, Duncan R. and {Zhang}, Bing},
        title = "{On the normalized FRB luminosity function}",
      journal = {\mnras},
         year = 2018,
        month = dec,
       volume = {481},
       number = {2},
        pages = {2320-2337},
          doi = {10.1093/mnras/sty2364},
archivePrefix = {arXiv},
       eprint = {1808.09929},
 primaryClass = {astro-ph.HE},
       adsurl = {https://ui.adsabs.harvard.edu/abs/2018MNRAS.481.2320L}
}

@ARTICLE{2023AJ....165..152M,
       author = {{Merryfield}, Marcus and {Tendulkar}, Shriharsh P. and {Shin}, Kaitlyn and {Andersen}, Bridget and {Josephy}, Alexander and {Good}, Deborah and {Dong}, Fengqiu Adam and {Masui}, Kiyoshi W. and {Lang}, Dustin and {M{\"u}nchmeyer}, Moritz and {Brar}, Charanjot and {Cassanelli}, Tomas and {Dobbs}, Matt and {Fonseca}, Emmanuel and {Kaspi}, Victoria M. and {Mena-Parra}, Juan and {Pleunis}, Ziggy and {Rafiei-Ravandi}, Masoud and {Sand}, Ketan R. and {Scholz}, Paul and {Smith}, Kendrick and {Stairs}, Ingrid H.},
        title = "{An Injection System for the CHIME/FRB Experiment}",
      journal = {\aj},
         year = 2023,
        month = apr,
       volume = {165},
       number = {4},
          eid = {152},
        pages = {152},
          doi = {10.3847/1538-3881/ac9ab5},
archivePrefix = {arXiv},
       eprint = {2206.14079},
 primaryClass = {astro-ph.IM},
       adsurl = {https://ui.adsabs.harvard.edu/abs/2023AJ....165..152M}
}

@ARTICLE{2021ApJS..257...59C,
       author = {{CHIME/FRB Collaboration} and {Amiri}, Mandana and {Andersen}, Bridget C. and {Bandura}, Kevin and {Berger}, Sabrina and {Bhardwaj}, Mohit and {Boyce}, Michelle M. and {Boyle}, P.~J. and {Brar}, Charanjot and {Breitman}, Daniela and {Cassanelli}, Tomas and {Chawla}, Pragya and {Chen}, Tianyue and {Cliche}, J.-F. and {Cook}, Amanda and {Cubranic}, Davor and {Curtin}, Alice P. and {Deng}, Meiling and {Dobbs}, Matt and {Dong}, Fengqiu Adam and {Eadie}, Gwendolyn and {Fandino}, Mateus and {Fonseca}, Emmanuel and {Gaensler}, B.~M. and {Giri}, Utkarsh and {Good}, Deborah C. and {Halpern}, Mark and {Hill}, Alex S. and {Hinshaw}, Gary and {Josephy}, Alexander and {Kaczmarek}, Jane F. and {Kader}, Zarif and {Kania}, Joseph W. and {Kaspi}, Victoria M. and {Landecker}, T.~L. and {Lang}, Dustin and {Leung}, Calvin and {Li}, Dongzi and {Lin}, Hsiu-Hsien and {Masui}, Kiyoshi W. and {McKinven}, Ryan and {Mena-Parra}, Juan and {Merryfield}, Marcus and {Meyers}, Bradley W. and {Michilli}, Daniele and {Milutinovic}, Nikola and {Mirhosseini}, Arash and {M{\"u}nchmeyer}, Moritz and {Naidu}, Arun and {Newburgh}, Laura and {Ng}, Cherry and {Patel}, Chitrang and {Pen}, Ue-Li and {Petroff}, Emily and {Pinsonneault-Marotte}, Tristan and {Pleunis}, Ziggy and {Rafiei-Ravandi}, Masoud and {Rahman}, Mubdi and {Ransom}, Scott M. and {Renard}, Andre and {Sanghavi}, Pranav and {Scholz}, Paul and {Shaw}, J. Richard and {Shin}, Kaitlyn and {Siegel}, Seth R. and {Sikora}, Andrew E. and {Singh}, Saurabh and {Smith}, Kendrick M. and {Stairs}, Ingrid and {Tan}, Chia Min and {Tendulkar}, S.~P. and {Vanderlinde}, Keith and {Wang}, Haochen and {Wulf}, Dallas and {Zwaniga}, A.~V.},
        title = "{The First CHIME/FRB Fast Radio Burst Catalog}",
      journal = {\apjs},
         year = 2021,
        month = dec,
       volume = {257},
       number = {2},
          eid = {59},
        pages = {59},
          doi = {10.3847/1538-4365/ac33ab},
archivePrefix = {arXiv},
       eprint = {2106.04352},
 primaryClass = {astro-ph.HE},
       adsurl = {https://ui.adsabs.harvard.edu/abs/2021ApJS..257...59C}
}

@ARTICLE{zhang2026,
       author = {{Zhang}, Zi-Liang and {Zhang}, Bing},
        title = "{Evidence for a Delayed Progenitor Population for CHIME non-repeating Fast Radio Bursts from a Self-Consistent Forward and Backward Inference Framework}",
      journal = {arXiv e-prints},
         year = 2026,
        month = jul,
          eid = {arXiv:2607.04792},
        pages = {arXiv:2607.04792},
          doi = {10.48550/arXiv.2607.04792},
archivePrefix = {arXiv},
       eprint = {2607.04792},
 primaryClass = {astro-ph.HE},
       adsurl = {https://ui.adsabs.harvard.edu/abs/2026arXiv260704792Z}
}

@article{Mas-Ribas:2026uhy,
    author = "Mas-Ribas, Lluis",
    title = "{Fast Radio Bursts probe Galaxy Evolution: Evidence and implications of a redshift-dependent FRB host DM}",
    eprint = "2606.25214",
    archivePrefix = "arXiv",
    primaryClass = "astro-ph.GA",
    month = "6",
    year = "2026"
}
\bibliographystyle{aasjournalv7}

\clearpage

\begin{table}
    \centering
    \renewcommand{\arraystretch}{1.3}
    \begin{tabular}{l c c c c c}
        \toprule
        \multicolumn{2}{c}{} & \multicolumn{2}{c}{\textbf{Catalog 1}} & \multicolumn{2}{c}{\textbf{Catalog 2}} \\
        \cmidrule(lr){3-4} \cmidrule(lr){5-6}
        \textbf{Parameter} & \textbf{Uniform prior range}      & DM and S/N                & Only DM     & DM and S/N           & Only DM \\
        \midrule
        \(\log \Phi_0\)               & \([-2, 20]\)           & \(11.33^{+0.55}_{-0.59}\)      & -                             & \(11.03^{+0.51}_{-0.53}\) & -                         \\
        \(\gamma\)                    & \([-2.50, 2.00]\)      & \(-1.54^{+0.24}_{-0.23}\)      & -                             & \(-1.72^{+0.15}_{-0.11}\) & -                         \\
        \(\log_{10} E_{\text{char}}\) & \([38.00, 49.00]\)     & \(41.46^{+0.26}_{-0.27}\)      & -                             & \(41.92^{+0.20}_{-0.18}\) & -                         \\
        \(\alpha\)                    & \([-5.00, 5.00]\)      & \(-0.88^{+0.47}_{-0.51}\)      & -                             & \(-0.87^{+0.59}_{-0.41}\) & -                         \\
        \(t_d\)                       & \([0, 13]\)            & \(0.75^{+0.56}_{-0.72}\)      & \(3.90^{+0.50}_{-0.41}\)      & \(1.68^{+0.75}_{-1.54}\) & \(3.53^{+0.11}_{-0.10}\)  \\
        \(\sigma_{t_d}\)              & \([0, 2]\)             & \(1.93^{+0.06}_{-1.73}\)      & \(1.99^{+0.01}_{-0.24}\)      & \(1.93^{+0.06}_{-0.87}\) & \(1.82^{+0.15}_{-0.14}\)  \\
        \(\ln \mu_{\text{host}}\)    & \([2.30, 6.30]\)       & \(4.10^{+0.23}_{-1.52}\)       & \(3.02^{+0.48}_{-0.03}\)      & \(4.56^{+0.29}_{-0.32}\) & \(3.70^{+0.10}_{-0.12}\)  \\
        \(\sigma_{\text{host}}\)      & \([0, 3]\)             & \(1.02^{+0.78}_{-0.48}\)       & \(1.98^{+0.02}_{-1.07}\)      & \(0.59^{+0.15}_{-0.20}\) & \(0.01^{+0.19}_{-0.01}\)  \\
        \bottomrule
    \end{tabular}
    \caption{The best-fit parameters for Catalogs 1 and 2. The first column lists the parameter names, the second column gives the adopted prior ranges, and the third through sixth columns correspond to results for different samples under different scenarios. For ease of visualization, the posterior distribution of some parameters are transformed from log space to linear space.}
    \label{tab}
\end{table}


\end{document}